\documentclass[useAMS,usedcolumn,usenatbib]{mn2e}
\usepackage{graphicx,color}
\usepackage{url}
\usepackage[normalem]{ulem} 

\usepackage{times}

\newcommand{\alm}{a_{\ell m}}
\newcommand*{\pseudoalm}[2]{\tilde{a}_{#1 #2}}
\newcommand{\Cee}{\mathcal{C}}
\newcommand{\Cl}{\Cee_\ell}

\newcommand{\Ctheory}{C^{\rmn{th}}}
\newcommand{\Cltheory}{\Ctheory_\ell}
\newcommand{\Ccorr}{\Cee} 

\newcommand{\ellmax}{\ell_{\rmn{max}}}
\newcommand{\fsky}{f_{\rmn{sky}}}

\newcommand*{\expectation}[1]{\langle #1\rangle}

\newcommand{\muK}{\rmn{\umu K}}

\newcommand{\healpix}{\textsc{healpix}}

\newcommand*{\satellite}[1]{\textit{#1}}

\newcommand{\WMAP}{\satellite{WMAP}}
\newcommand{\Planck}{\satellite{Planck}}

\newcommand*{\Planckmap}[1]{\texttt{#1}}
\newcommand{\smica}{\Planckmap{SMICA}}
\newcommand{\nilc}{\Planckmap{NILC}}
\newcommand{\sevem}{\Planckmap{SEVEM}}
\newcommand{\CR}{\Planckmap{Commander-Ruler}}

\newcommand{\LCDM}{$\Lambda$CDM}

\newcommand*{\unit}[1]{\;\rmn{#1}}
\renewcommand*{\vec}[1]{\bmath{#1}}
\newcommand*{\unitvec}[1]{\vec{\hat{#1}}}
\newcommand{\mat}[1]{\mathbfss{#1}}

\newcommand{\Cinv}{\mat C^{-1}}

\newcommand*{\WignerthreeJ}[6]{\left(
\begin{array}{ccc} #1 & #2 & #3 \\ #4 & #5 & #6 \end{array}
\right)}
\newcommand*{\sci}[1]{\times 10^{#1}}
\newcommand{\trace}{\mathop{\mathrm{Tr}}\nolimits}
\newcommand*{\group}[1]{#1}

\newcommand{\Shalf}{S_{1/2}}

\newcommand{\iimag}{\rmn{i}}
\newcommand{\eexp}{\rmn{e}}
\newcommand{\dderiv}{\rmn{d}}

\title[Alignments from WMAP and Planck]{Large-scale alignments from WMAP
  and Planck}
\author[C.J. Copi, D. Huterer, D.J. Schwarz and G.D. Starkman]
{Craig J. Copi$^{1}$\thanks{E-mail: cjc5@cwru.edu},
 Dragan Huterer$^{2}$\thanks{E-mail: huterer@umich.edu},
 Dominik J. Schwarz$^{3}$\thanks{E-mail: dschwarz@physik.uni-bielefeld.de}
 and 
 Glenn D. Starkman$^{1,4}$\thanks{E-mail: glenn.starkman@case.edu}\\
 $^{1}$CERCA/Department of Physics/ISO, Case Western Reserve University,
 Cleveland, OH 44106-7079, USA\\
 $^{2}$Department of Physics, University of Michigan, 
 450 Church St, Ann Arbor, MI 48109-1040, USA\\
 $^{3}$Fakult\"at f\"ur Physik, Universit\"at Bielefeld,
 Postfach 100131, 33501 Bielefeld, Germany\\
 $^{4}$Physics Department, Theory Unit, CERN,
 CH-1211 Gen\`eve 23, Switzerland}

\begin{document}

\date{Accepted xxxx. Received xxxx; in original form xxxx}

\pagerange{\pageref{firstpage}--\pageref{lastpage}} \pubyear{2013}

\maketitle

\label{firstpage}

\begin{abstract}
  We revisit the alignments of the largest structures observed in the
  cosmic microwave background (CMB) using the seven and nine-year
  \WMAP\ and first-year \Planck\ data releases.  The observed alignments --
  the quadrupole with the octopole and their joint alignment with the
  direction of our motion with respect to the CMB (the dipole direction)
  and the geometry of the Solar System (defined by the Ecliptic plane) --
  are generally in good agreement with results from the previous
  \WMAP\ data releases.  However, a closer look at full-sky data on the
  largest scales reveals discrepancies between the earlier \WMAP\ data
  releases (three to seven-year) and the final, nine-year release. There
  are also discrepancies between all the \WMAP\ data releases and the
  first-year \Planck\ release. Nevertheless, both the \WMAP\ and
  \Planck\ data confirm the alignments of the largest observable CMB modes
  in the Universe.  In particular, the $p$-values for the mutual alignment
  between the quadrupole and octopole, and the alignment of the plane
  defined by the two with the dipole direction, are both at the greater
  than $3$-sigma level for all three \Planck\ maps studied. We also
  calculate conditional statistics on the various alignments and find that
  it is currently difficult to unambiguously identify a leading anomaly
  that causes the others or even to distinguish correlation from causation.
\end{abstract}


\section{Introduction}

Cosmic microwave background (CMB) maps produced by the \WMAP\ and, most
recently, the \Planck\ collaborations provide an unprecedented view into the
physics of the early Universe.  These data help constrain the parameters of
the standard cosmological model, \LCDM, to the per cent level accuracy.  They
also point out some intriguing anomalies, particularly on the largest
angular scales or at the lowest multipole moments.  One anomaly, and
the one of interest for this work, is alignments. The quadrupole and
octopole are found to be mutually aligned and they define axes that are
unusually perpendicular to the Ecliptic pole and parallel to the direction
of our motion with respect to the rest frame of the CMB (the dipole
direction).  Another anomaly, the lack of correlations on large angular
scales, is the subject of a companion work \citep{CHSS-Planck-R1-ctheta}.
For reviews of the large-angle anomalies \emph{before} the final \WMAP\ and
the first cosmological \Planck\ data releases, see \citet{WMAP7-anomalies}
and \citet{CHSS-review}.

The study of alignments requires precise measurements of the full-sky
CMB\@.  Conventional wisdom tells us that the largest scales in the
Universe should be well measured. The basic argument is straightforward:
measurements by \WMAP\ and \Planck\ are signal dominated on small angular
scales.  Large numbers of these measurements are averaged to determine the
large-scale structure of the CMB which further reduces the noise leading to
very precise determinations.  Thus, even though cosmic variance is large
and our Universe could have been drawn from a broadly distributed ensemble,
our particular realization can be very well measured.  Both \WMAP\ and
\Planck\ are measuring the same CMB sky so should agree very well on the
largest scales.

Unfortunately, the real-world is more complicated.  Though the statistical
noise is small, residual contamination due to foregrounds, even after cleaning
the maps, remains a source of significant uncertainty.  These issues are of
paramount importance in the study of alignments.  In the nine-year data
release the \WMAP\ team states: `We conclude that our ability to remove
foregrounds is the limiting factor in measurement of the cosmological
quadrupole+octopole alignment' \citep{WMAP9-results}.  They then argue that
the statistical significance of the alignment is weakened due to this.  In a
similar context the \Planck\ team states: `Residual foregrounds (mostly on the
Galactic plane) present in the four \Planck\ CMB estimates could influence the
reconstruction of the low-order multipoles'
\citep{Planck-R1-XXIII}.
(Note that their four maps include the \CR\ map in addition to the three we
discuss below.)  They proceed to Wiener filter their maps to further reduce
contamination which has a small effect on alignments.  This observation is
not new: for example, \citet{Chiang2007} found signs of residual foregrounds
in the \WMAP\ three-year data release on these large scales.  Such residual
contaminations persist in the data to the present time.

The goal of this paper is to study these large-angle alignments with the
\Planck\ one-year data and to compare them to the \WMAP\ seven and
nine-year data.  The \Planck\ collaboration included a brief discussion of
alignments in their extensive study of isotropy in the CMB
\citep{Planck-R1-XXIII}. In their presentation they did not include results
from the principal mathematical tool employed here -- the multipole
vectors; moreover, we use a somewhat different approach to generate the
full-sky maps, and use a much larger set of Monte-Carlo simulations to
obtain the statistical inferences. This work therefore complements the
existing detailed study of other tests of isotropy, such as the
hemispherical power asymmetries and moments of the temperature field
\citep{Planck-R1-XXIII}.

To study the large-angle alignments it is necessary to define statistics
and assign significance to the data based on them.  In recent years the
trend has been towards applying a Bayesian analysis to all statistical
questions.  Broadly speaking, the Bayesian approach is designed to compare
models and for parameter estimation within a model.  Even the problem of
null hypothesis testing in the Bayesian approach is reduced to a model
comparison; that between the full model and a subset of the full model with
a restricted parameter set, some parameters fixed, etc.  When there is a
model with no serious competitors, such as in cosmology with \LCDM,
Bayesian statistics struggles to even ask the question of the consistency
between the model and the data (though see \citealt{Starkman:2008py} for a
possible, if computationally challenging, approach).  At the present time
in cosmology \emph{there are no compelling alternative models} that can
account for the anomalies.  Clever ideas have been proposed to explain some
of the anomalies (e.g.~\citet{Frisch2005, Gordon2005, Rakic2006a,
  Alnes2006, Inoue2006, Pullen2007, Dikarev2008, Ramirez2009, Peiris2010}),
but no model that explains all, or even most, of them exists.  It is not
even clear whether the origin of the anomalies is cosmological,
astrophysical foregrounds, systematic (instrumental, map making, etc.), or
simply statistical, although it could be argued that since the
\Planck\ satellite and data reduction is very different from that provided
by the \WMAP\ satellite, systematic effects are unlikely to explain the
existence of shared anomalies.  Due to these issues, we adopt the
frequentist approach consistent with that used in previous work (see
\citealt{CHSS-anomalies}, for example).  The frequentist approach is well
suited for this specific problem -- to address the question of tension
between the model and the data, and, if there is one, where this tension
lies.  This allows the data to point the way toward the source of any
potential discrepancy independent of finding a better model to describe it.

This paper is organized as follows. In Sec.~\ref{secn:large-scales} we take a
first look at the large-angle \Planck\ data and compare the angular power
spectrum at the lowest multipoles between \Planck\ and \WMAP\ releases. In
Sec.~\ref{secn:fullsky} we describe the methodology of how we arrive at an
ensemble of full-sky maps using harmonic inpainting, and how we correct for
the effect of our motion through the CMB rest frame on the quadrupole. In
Sec.~\ref{secn:alignment} we describe the statistics that we use (and that we
developed previously for our \WMAP\ analyses), and in Sec.~\ref{secn:results}
we carry out a frequentist analysis to quantify the various alignments.  We
conclude in Sec.~\ref{secn:Conclusions}.

\section{State of large-scale data}
\label{secn:large-scales}

\begin{table}
  \caption{The power spectrum coefficients, $D_\ell$, in units of $\muK^2$
    as reported by \WMAP\ and \Planck. All values are based on a maximum
    likelihood estimator.  Since the one-year \WMAP\ reported values were
    based on pseudo-$\Cl$ estimators, they have been excluded from this
    table. The $\Shalf$ values have been computed for $\ellmax=100$ unless
    otherwise stated.}
  \label{tab:large-scale-Dl}
  \begin{tabular}{ld{0}d{0}d{0}d{0}d{4.0}} \hline
    Data Release
    & \multicolumn{1}{c}{$D_2$}
    & \multicolumn{1}{c}{$D_3$}
    & \multicolumn{1}{c}{$D_4$}
    & \multicolumn{1}{c}{$D_5$}
    & \multicolumn{1}{c}{ $\Shalf\unit{(\muK^4)}$}
    \\ \hline
    \WMAP\ 3yr & 211 & 1041 & 731 & 1521 & 8330 \\
    \WMAP\ 5yr & 213 & 1039 & 674 & 1527 & 8915 \\
    \WMAP\ 7yr & 201 & 1051 & 694 & 1517 & 8938 \\
    \WMAP\ 9yr & 151 & 902 & 730 & 1468 & 5797 \\
    \Planck\ R1  & 299 & 1007 & 646 & 1284 & 8035\rlap{$^a$}
    \\
    \hline
  \end{tabular}
  \\ ${}^a$This $\Shalf$ has been calculated for $\ellmax=49$ since
  \Planck\ only provides binned values for $\ell\ge50$.
\end{table}

Since full-sky data is required for a study of alignments it is important
to understand its current state.  A high level view of the data can be
obtained through the power spectrum coefficients
\begin{equation}
  D_\ell \equiv \frac{\ell(\ell+1)}{2\upi} \Cl .
  \label{eq:Dl-definition}
\end{equation}
The results reported by \WMAP\ and \Planck\ are summarized in
Table~\ref{tab:large-scale-Dl}.\footnote{All CMB data is available from the
  Lambda site, \url{http://lambda.gsfc.nasa.gov/}, including links to both
  \WMAP\ and \Planck\ results.  The \Planck\ results may directly be
  obtained via the \Planck\ Legacy Archive,
  \url{http://archives.esac.esa.int/pla/}.}  In all cases these values are
given in $\muK^2$ and are based on a maximum likelihood estimator.  Since
the one-year \WMAP\ analysis employed a pseudo-$\Cl$ based estimator it has
not been included in the table.

The $\Shalf$ statistic defined in the one-year \WMAP\ data release
\citep{WMAP1-cosmology} is a convenient and discriminating tool to
quantify the lack of correlations on large angular scales.  The statistic can be
calculated as
\begin{equation}
  \Shalf \equiv \int_{-1}^{1/2} [\Ccorr(\theta)]^2 \dderiv(\cos\theta)
  = \sum_{\ell=2}^{\ellmax} \Cl I_{\ell\ell'} \Cee_{\ell'},
\end{equation}
where the $I_{\ell\ell'}$ are components of an easily calculated matrix.
For a more thorough discussion of the $\Shalf$ from the \Planck\ data see
\citet{CHSS-Planck-R1-ctheta}.  In Table~\ref{tab:large-scale-Dl} the $\Shalf$
has been calculated for $\ellmax=100$ except for \Planck\ which only provides
binned $\Cl$ for $\ell\ge 50$.

The noise only contribution to the uncertainty in the values in
Table~\ref{tab:large-scale-Dl} is estimated in \WMAP\ from the Fisher matrix
to be $\sigma_{D_\ell}\sim 10\unit{\muK^2}$ in all cases; typically slightly
larger in the earlier data releases and slightly smaller in the later ones.
This sets the scale for the expected statistical scatter in the data.  From
the table we see that the three-year through seven-year \WMAP\ data releases
are in good agreement and provide a consistent picture of the large-scale
Universe.  The differences are most likely due to systematic analysis
improvements such as beam shape determination, point source identification,
and masking.  Quite surprisingly the nine-year \WMAP\ data release provides a
markedly different view of the large-scale Universe and even more surprisingly
the first \Planck\ data release provides yet another different view.  In
particular, comparing the quadrupole, $D_2$, we see that the nine-year
\WMAP\ and first-year \Planck\ results differ by a factor of two, or roughly
$15$ times the approximate noise error!

\begin{table}
  \caption{The power spectrum coefficients, $D_\ell$, in units of $\muK^2$
    extracted from cleaned, full-sky maps provided by \WMAP\ and \Planck.}
  \label{tab:large-scale-maps-Dl}
  \begin{tabular}{ld{0}d{0}d{0}d{0}d{4.0}} \hline
    Map
    & \multicolumn{1}{c}{$D_2$}
    & \multicolumn{1}{c}{$D_3$}
    & \multicolumn{1}{c}{$D_4$}
    & \multicolumn{1}{c}{$D_5$}
    & \multicolumn{1}{c}{ $\Shalf\unit{(\muK^4)}$}
    \\ \hline
    \WMAP\ ILC 1yr  & 195 & 1053 & 834 & 1667 & 8190 \\
    \WMAP\ ILC 3yr  & 248 & 1051 & 756 & 1588 & 8476 \\
    \WMAP\ ILC 5yr  & 243 & 1052 & 730 & 1591 & 8642 \\
    \WMAP\ ILC 7yr  & 240 & 1048 & 731 & 1593 & 8528 \\
    \WMAP\ ILC 9yr  & 243 & 1013 & 709 & 1612 & 8156 \\
    \Planck\ \nilc  & 209 & 863 & 704 & 1379 & 4816 \\
    \Planck\ \sevem & 205 & 798 & 736 & 1207 & 3766 \\
    \Planck\ \smica & 239 & 925 & 713 & 1494 & 6309\\
    \hline
  \end{tabular}
\end{table}

An alternative view of the large-scale Universe comes from the foreground
cleaned, full-sky maps provided in the data releases (we restrict our
analysis to the maps officially released by the \WMAP\ and
\Planck\ collaborations).  The $D_\ell$ extracted from these maps are shown
in Table~\ref{tab:large-scale-maps-Dl}. For \WMAP\ the maps are cleaned
using the ILC method \citep{Eriksen2004-LILC,WMAP3-temperature}.  For
\Planck\ multiple cleaning procedures were used and three, full resolution
maps based on them, the \smica, \nilc, and \sevem\ maps, have been provided
\citep{Planck-R1-XII}.  Further, for the \smica\ and \nilc\ maps a small
\textit{inpainting mask} with $\fsky\approx0.97$ has been defined by the
\Planck\ team.  This small region has been inpainted using a constrained
realization and it is these inpainted maps that were analysed for the
values provided in the table.  From these results we see that the
\WMAP\ full-sky maps from all data releases are in reasonably good mutual
agreement.  The one-year map is somewhat discrepant and the nine-year map
is also somewhat different in its $\Shalf$ value, but otherwise the
cleaning procedure employed throughout the years has produced relatively
stable results. For the \Planck\ maps this is not the case.  Though the
cleaning procedures do produce quadrupoles more in line with the one from
\WMAP, there is still a large discrepancy between the large-scale structure
internally among the \Planck\ maps and in comparison with \WMAP\@.
Moreover, these three maps were not the only cleaned maps produced, just
the best three (four including also the \CR\ method), selected by the
\Planck\ team, based on unpublished criteria.  The larger spread of results
from the \Planck\ maps when compared to the \WMAP\ ILC maps of different
years does not come as a surprise. While the \WMAP\ collaboration published
similar incarnations of the same ILC method over the years, the three
analysed \Planck\ maps are based on (radically) different assumptions about
the foreground.

Naturally many details have been ignored in these comparisons.  For
example, a visual examination of the \sevem\ map shows clear signs of small
scale contaminations near the Galactic plane.  This region is inpainted in
the \smica\ and \nilc\ maps and suggests we should be careful interpreting
results based on the \sevem\ map.  Also, for all of the maps larger masks,
called \textit{validity masks}, have been defined by the \Planck\ team to
define a more conservative (i.e.\ smaller) set of pixels that are believed
to be clean. These masks have sky fractions of $\fsky=0.88$ for the
\smica\ map, $\fsky=0.93$ for the \nilc\ map, and $\fsky=0.76$ for the
\sevem\ map. As explained below, we will mostly use the inpainting masks,
except for the \sevem\ map for which we will use the validity mask.

\section{Full-sky maps}\label{secn:fullsky}

Perhaps contrary to our conventional wisdom, the largest scales in the CMB
have been measured precisely but not (yet) necessarily accurately; this is
immediately clear by inspecting the angular power as represented in
Tables~\ref{tab:large-scale-Dl} and
\ref{tab:large-scale-maps-Dl}. Therefore, care must be taken when analysing
full-sky data and when interpreting the results. In this section we
describe our methodology for arriving at an ensemble of maps that represent
the \Planck\ full-sky. The two particular techniques that we will apply are
inpainting and Doppler correction for the quadrupole.

For the purpose of studying alignments in \LCDM\ cosmologies, all that
matters is the phase structure of the $\alm$ -- i.e. the relative values of
$\alm$ of equal $\ell$ and all $m$.  The statistics used to study
alignments are constructed so as to be independent of the power spectrum.
Thus any power spectrum could be used to produce realizations for studying
the distributions of these statistics, provided the $\alm$ have random
phases, and the underlying distribution is statistically isotropic.
Despite this, some of the discussions below will refer to the magnitudes of
the $\alm$.  To be consistent with our companion analysis of the two-point
angular correlation function \citep{CHSS-Planck-R1-ctheta} we use the
\Planck\ base best-fitting power spectrum.  This power spectrum is freely
available along with all the \Planck\ data and is based on the parameters
listed in the first column of table 2 in \citet{Planck-R1-XVI}.

\subsection{Inpainting}

One approach to handling particularly contaminated regions of a map is to
excise those regions and to replace them with synthetic data constrained to
agree with the uncontaminated regions of the map.  In general, masking the
skies introduces coupling among the modes so this must be used with some
care.  The relationship between the uncontaminated partial map and a
representation of the full-sky information (either as a map or spherical
harmonic coefficients, $\alm$) is an under-determined linear system.  Due
to this there is not a unique solution and extra information is required to
re-fill the contaminated regions.  One set of techniques to achieve this
re-filling, that is the extra assumptions imposed and the balance between
these assumptions and the type and quality of fit to the uncontaminated
regions, is known as inpainting.  Inpainting has been studied in numerous
contexts and many approaches have been explored for CMB maps (see
\citealt{Inoue_harmonic_inpaint} and especially \citealt{Starck2013} for a
discussion of techniques in the context of low-$\ell$ reconstruction). The
\Planck\ analysis has implemented a particular form of inpainting
\citep{Benoit-Levy2013,Planck-R1-XII} used in the \smica\ and \nilc\ maps
discussed above. For other related work on the topic, see
\citet{Abrial_inpaint,Dupe2011,Bucher_inpaint,Nishizawa_inpaint}.

Inpainting is not a simple cure for contamination.  The resulting full-sky
information will depend on the extra assumptions (type of inpainting
performed) and also on the mask employed.  Choosing a large mask will
remove the most contamination but will also lead to a large variance, poor
determination, of the full-sky information.  In this work, our goal is to
study the large-scale anomalies present in the full-sky data independent of
their origin. If we were interested in only a cosmological origin for
alignments, we would restrict the analysis to the cleanest portions of the
sky by employing a large mask.  A large mask washes out the results making
definitive statements about the alignments impossible.  On the other hand,
employing a small mask allows for definitive statements about the presence
of alignments but reduces the ability to determine their origin.  Since our
focus is on the existence of alignments in the data as provided, our main
results will be based on inpainting using the aforementioned smaller
inpainting masks with $\fsky\approx 0.97$.

\subsection{Harmonic inpainting}
\label{ssecn:harmonic_inpainting}

Harmonic inpainting as applied to the
CMB~\citep{Inoue_harmonic_inpaint,Inpainting2012} is an application of
constrained Gaussian realizations for reconstructing the $\alm$ from a
masked sky.  It \emph{assumes} the usual statistical properties,
Gaussianity and isotropy, but also produces probability distributions which
can be used in subsequent analyses.  Here we mainly follow
\citet{Inpainting2012} but have changed the notation slightly.  The
technique is intimately related to pseudo-$\alm$ reconstruction as will be
evident in what follows.

\subsubsection{Algorithm}

Consider a temperature map, $T(\unitvec e)$, and a mask, $W(\unitvec e)$.
Here, and in what follows, $\unitvec e$ represents the usual radial unit
vector that in Cartesian coordinates may be written as
\begin{equation}
  \unitvec e = (\sin\theta\cos\phi, \sin\theta\sin\phi, \cos\theta).
\end{equation}
These quantities may be expanded in spherical harmonics as
\begin{equation}
  T(\unitvec e) = \sum_{\ell' m'} a_{\ell' m'} Y_{\ell' m'}(\unitvec e),
  \quad W(\unitvec e) = \sum_{\tilde\ell \tilde m} w_{\tilde\ell \tilde m}
  Y_{\tilde\ell \tilde m}(\unitvec e).
\end{equation}
We may also expand the masked sky in terms of the pseudo-$\alm$,
denoted by $\pseudoalm{\ell}{m}$, and relate this to the
previous expansions by
\begin{eqnarray}
  W(\unitvec e) T(\unitvec e)
  & = & \sum_{\ell m} \pseudoalm{\ell}{m} Y_{\ell m}(\unitvec e) \\
  & = & \sum_{\tilde\ell \tilde m} w_{\tilde\ell \tilde m}
  Y_{\tilde\ell \tilde m}(\unitvec e)
  \sum_{\ell' m'} a_{\ell' m'} Y_{\ell' m'}(\unitvec e). \nonumber
\end{eqnarray}
This is a standard procedure: we solve for the pseudo-$\alm$ by integrating
and using the properties of the spherical harmonics.  This leads to an
integral over three spherical harmonics which may be evaluated in terms of
the Gaunt coefficients and most easily represented by the Wigner $3J$
symbols.  The end result is
\begin{equation}
  \pseudoalm{\ell}{m} = \sum_{\ell' m'} F_{\ell m;\ell' m'} a_{\ell' m'} ,
\end{equation}
where
\begin{eqnarray}
  F_{\ell m;\ell' m'}
  & \equiv &
  (-1)^m \sum_{\tilde\ell \tilde m} w_{\tilde\ell \tilde m}
  \sqrt{\frac{(2\ell+1)(2\ell'+1)(2\tilde\ell+1)}{4\upi}} \nonumber \\
  & & {} \quad \times
  \WignerthreeJ{\ell'}{\tilde\ell}{\ell}{0}{0}{0}
  \WignerthreeJ{\ell'}{\tilde\ell}{\ell}{m'}{\tilde m}{-m}.
  \label{eq:coupling-F}
\end{eqnarray}
More compactly we write this in matrix notation as
\begin{equation}
  \vec{\tilde a} = \mat F \vec a.
\end{equation}
Moreover, we define the following matrices which are simplified using the
statistical isotropy of the underlying $\alm$,
\begin{equation}
  \mat b \equiv \expectation{\vec a\vec{\tilde a}^\dagger}
  =
  \mat{C}_a \mat F^\dagger, \quad \mat C \equiv \expectation{\vec{\tilde
      a}\vec{\tilde a}^\dagger} = \mat F \mat{C}_a \mat F^\dagger,
\end{equation}
where the angle brackets represent an ensemble average and $\mat{C}_a$
represents the diagonal matrix of the $\Cltheory$ from the best-fitting
\LCDM\ model, $\mat{C}_a\equiv\expectation{\vec a \vec a^\dagger}$

For a given map and mask we calculate $\mat F$ and $\vec{\tilde a}$.  From
these harmonic inpainting proceeds by generating realizations of \LCDM\ as
$\alm^{\rmn{mc}}$, that is, the $\alm^{\rmn{mc}}$ are Gaussian random
variables with $|\alm^{\rmn{mc}}|\in N(0, \Cltheory)$ and random phases.
These are then corrected giving us constrained realizations as
\begin{equation}
  \vec a^{\rmn{inp}} = \vec a^{\rmn{mc}} + \mat b \Cinv (\vec{\tilde a} - \mat F
  \vec a^{\rmn{mc}}).
\end{equation}
We now explain how to efficiently calculate the matrix $\Cinv$.

\begin{figure}
  \includegraphics[width=3.5in]{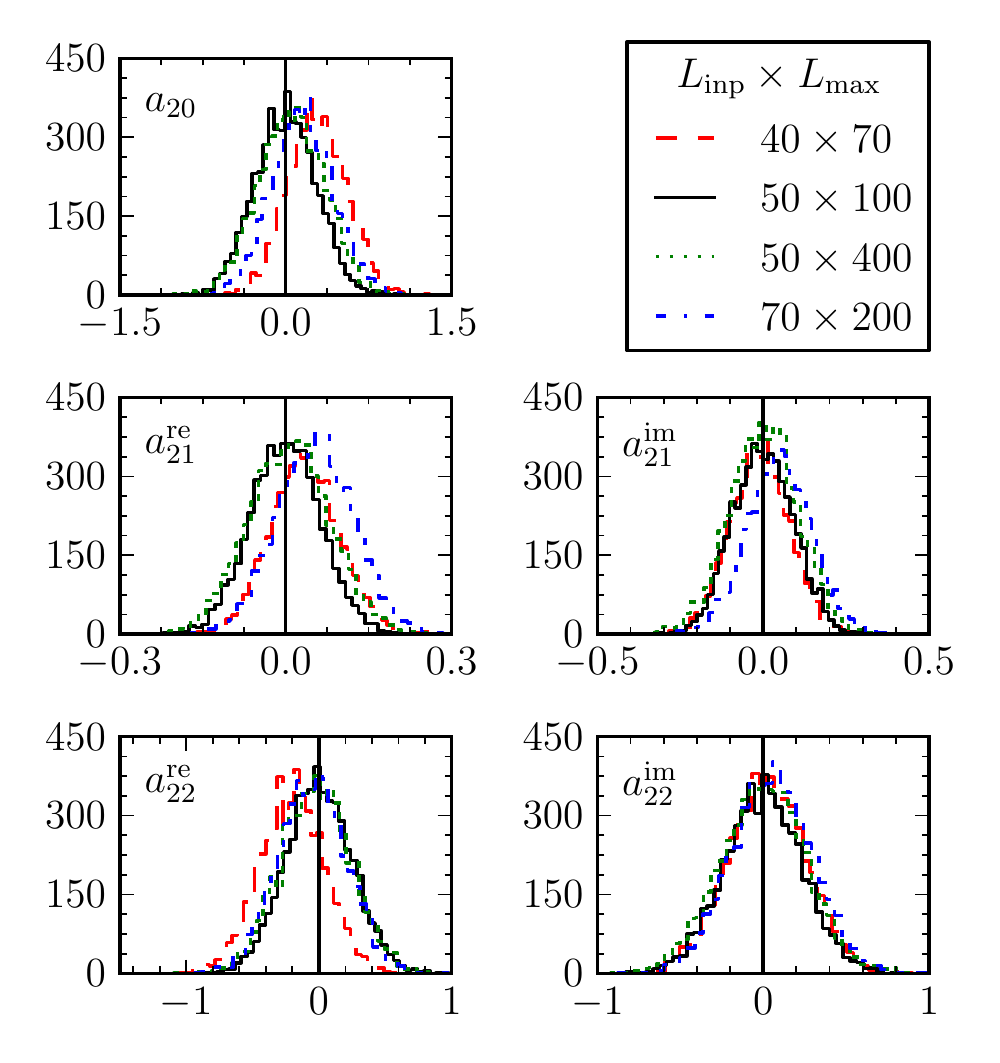}
  \caption{Histograms of the $a_{2m}$ in $\muK$ from $5000$ harmonic
    inpaintings of the \smica\ map using its inpainting mask for various
    choices of scales $L_{\rmn{inp}}\times L_{\rmn{max}}$.  The histograms
    have been shifted by the average from the $50\times100$ case so are
    centred roughly around zero (and precisely on zero for the black
    curve).  These histograms show little sensitivity in the choice of
    $L_{\rmn{inp}}\times L_{\rmn{max}}$ and that the $a_{2m}$ are well
    reconstructed, typically to within about $1\unit{\muK}$.}
  \label{fig:a2m-smica-L1-L2}
\end{figure}

\begin{figure}
  \includegraphics[width=3.5in]{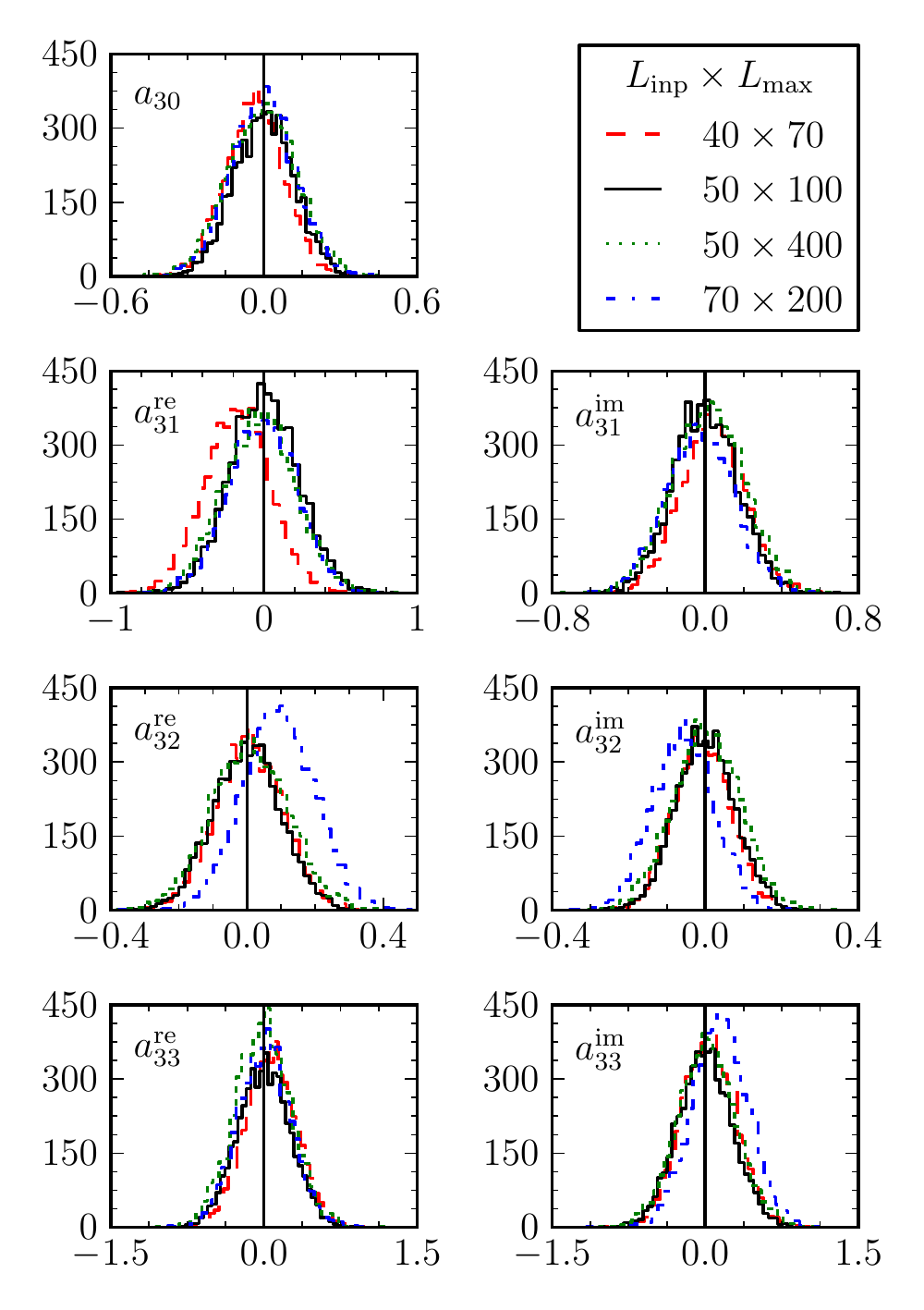}
  \caption{Same as Fig.~\ref{fig:a2m-smica-L1-L2}, now for the $a_{3m}$.}
  \label{fig:a3m-smica-L1-L2}
\end{figure}

\subsubsection{Inversion and marginalization}

Calculating the matrix $\Cinv$ is complicated by the fact that the
underlying map which we are trying to inpaint may contain a monopole and
dipole (on the full- or cut-sky).  We wish to remove these contributions.
This can be done in a number of ways; we will marginalize over them using a
method based on that of \citet{Slosar2004} with some modifications.

To begin we construct the rectangular coupling matrix, $\mat F$, as above
using the best-fitting \LCDM\ $\Cltheory$ with
$\Ctheory_0=\Ctheory_1=0$. Let $\mat f$ represent the columns of $\mat F$
corresponding to the monopole and dipole (typically the first four columns)
and $\hat{\mat F}$ be the remainder of the columns.  With this we may
marginalize over the monopole and dipole in the correlation matrix by first
writing it as
\begin{equation}
  \mat C = 
  \hat{\mat F}\mat{C}_a\hat{\mat F}^\dagger + \lambda \mat f {}
  \mat f^\dagger
  \equiv 
  \hat\mat{C} + \lambda \mat f {} \mat f^\dagger,
\end{equation}
where $\lambda$ is a large parameter, preferably
$\lambda\rightarrow\infty$.  This can be accomplished analytically when
calculating the inverse using a result from \citet{Rybicki1992},
\begin{eqnarray}
  \Cinv
  & = &
  \lim_{\lambda\rightarrow\infty} \left(\hat{\mat C} + \lambda \mat f {}\mat
  f^\dagger \right)^{-1} \nonumber \\
  & = &
  \lim_{\lambda\rightarrow\infty} \left[ \hat{\mat C}^{-1}
    - \hat{\mat C}^{-1} \mat f \left( \lambda^{-1} + \mat f^\dagger
    \hat{\mat C}^{-1} \mat f \right)^{-1} \mat f^\dagger
    \hat{\mat C}^{-1} \right] \\
  & = & \hat{\mat C}^{-1} - \hat{\mat C}^{-1} \mat f \left( \mat f^\dagger
  \hat{\mat C}^{-1} \mat f \right)^{-1} \mat f^\dagger
  \hat{\mat C}^{-1}. \nonumber
\end{eqnarray}
This last line is now easy to calculate.  Since $\hat{\mat C}$ is a
correlation matrix it is Hermitian and positive definite so may be
efficiently inverted using a Cholesky decomposition.  We have an extra
matrix we need to invert, the one in parenthesis in the last line of the
previous equation.  However, this is a $4\times4$ correlation matrix so is
also easy to handle.

Initially, adoption of harmonic inpainting seems risky since it imposes the
assumptions of Gaussianity and statistical isotropy on the full-sky $\alm$.
However, this is outweighed by clear advantages: the inpainting generates
statistical distributions of the full-sky $\alm$ thus allowing accurate
statements for any test that rests on the full CMB sky, and it is
straightforward to implement.  Crucially, since we have restricted our main
analysis to masks with $\fsky\approx 0.97$, the results will be insensitive
to the inpainting assumptions.

\begin{figure}
  \includegraphics[width=3.5in]{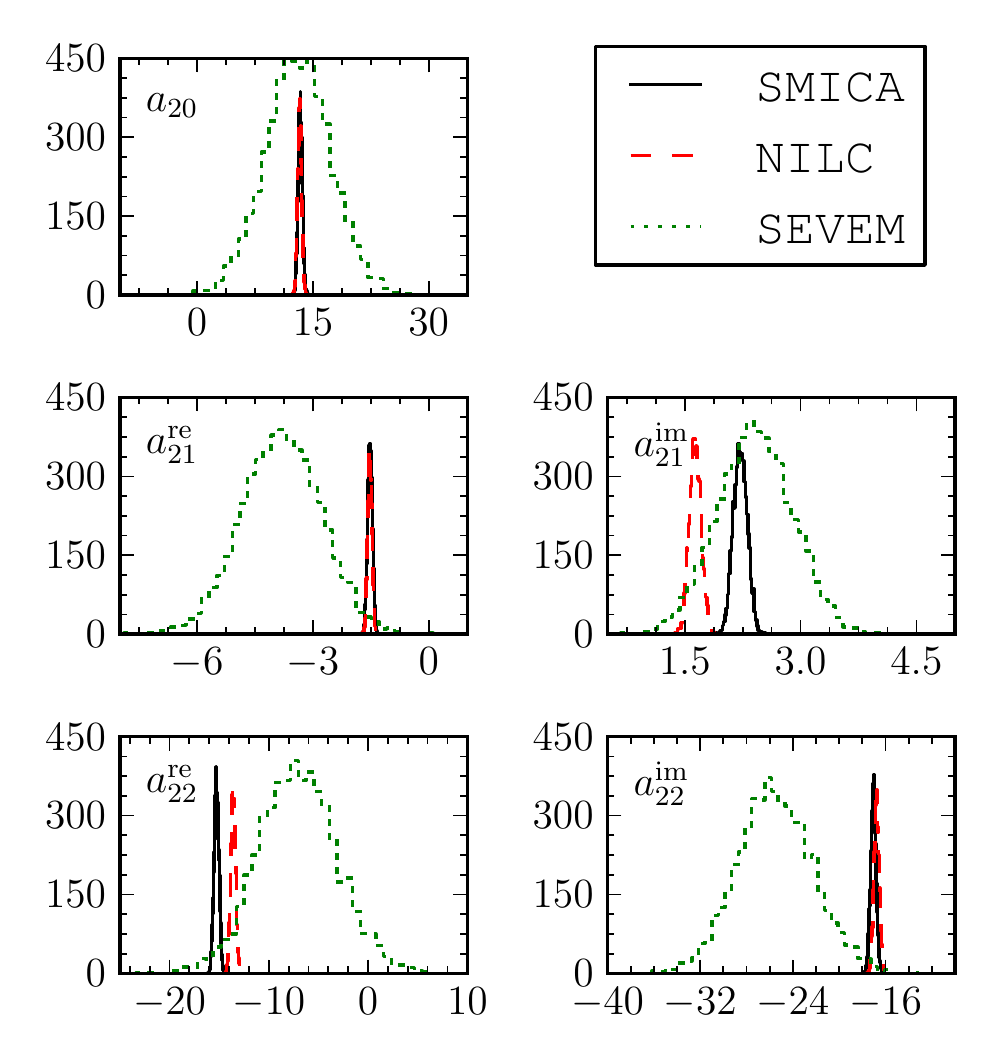}
  \caption{Histograms of the $a_{2m}$ in $\muK$ from $5000$ harmonic
    inpainting of the \Planck\ maps for the scales $L_{\rmn{inp}}\times
    L_{\rmn{max}}=50\times100$.  For the \smica\ and \nilc\ maps the areas
    covered by their inpainting masks with $\fsky=0.97$ have been
    inpainted.  For the \sevem\ map its validity mask with $\fsky=0.76$ has
    been used.  From these histograms we see that the $a_{2m}$ are mostly
    consistent among the maps, see the text for details.}
  \label{fig:a2m-maps-50-100}
\end{figure}

\begin{figure}
  \includegraphics[width=3.5in]{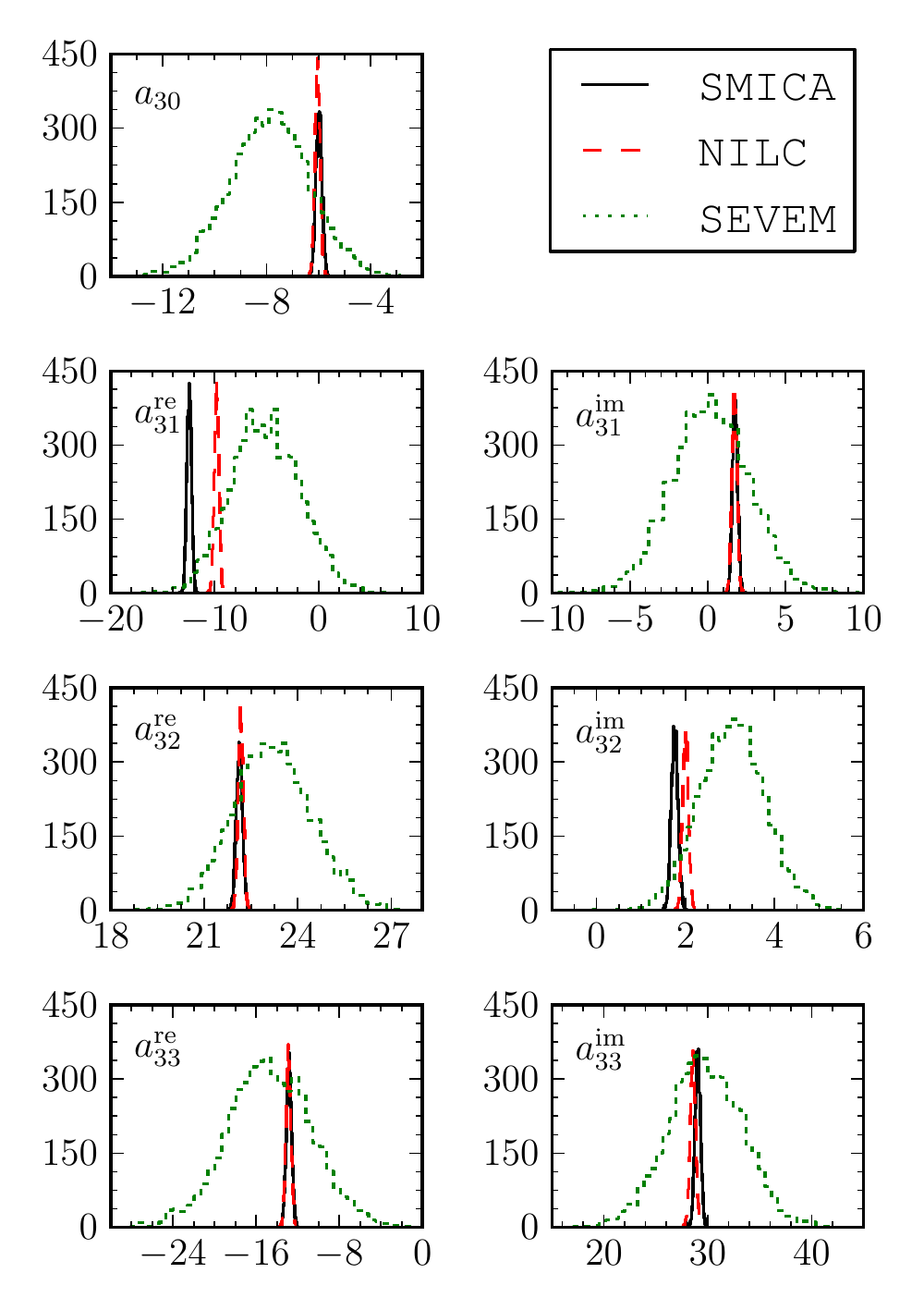}
  \caption{Same as Fig.~\ref{fig:a2m-maps-50-100}, now for the $a_{3m}$.}
  \label{fig:a3m-maps-50-100}
\end{figure}

\subsubsection{Results}

Two scales appear in harmonic inpainting: we denote by $L_{\rmn{inp}}$ the
maximum multipole reconstructed by the inpainting and by $L_{\rmn{max}}$
the maximum multipole used in the reconstruction (i.e.\ the largest
coupling multipole).  In other words, the coupling matrix $\mat F$ defined
in Eq.~(\ref{eq:coupling-F}) is a rectangular matrix of size
$(L_{\rmn{inp}}+1)^2\times(L_{\rmn{max}}+1)^2$.  For the work performed
here we are only interested in the quadrupole and octopole and (mainly) in
the \smica\ and \nilc\ maps with their inpainting masks with
$\fsky\approx0.97$.  We study the sensitivity of harmonic inpainting with
these constraints in mind.

To study the sensitivity to choices of $L_{\rmn{inp}}$ and $L_{\rmn{max}}$
we begin with the \smica\ map.  Figs.~\ref{fig:a2m-smica-L1-L2} and
\ref{fig:a3m-smica-L1-L2} show histograms from $5000$ reconstructions for
various choices of $L_{\rmn{inp}}\times L_{\rmn{max}}$.  We have chosen the
$L_{\rmn{inp}}\times L_{\rmn{max}}=50\times 100$ case as the standard.  The
histograms in these figures are shifted by the average value from this
($50\times 100$) case.  The $x$-axis in the histograms is in units of
$\muK$ and shows that the $\alm$ are reconstructed to within about
$1\unit{\muK}$ in all cases.  This is not surprising: given the small
fraction of the sky inpainted it is expected that $\mat F$ is diagonally
dominant; the resulting inpainting should not be sensitive to the choice of
scales (provided they are large enough) and the $\alm$ should be well
determined.

With the choice $L_{\rmn{inp}}\times L_{\rmn{max}}=50\times100$ we next
compare inpainting of the various \Planck\ maps.  These histograms based on
$5000$ reconstructions are shown in Figs.~\ref{fig:a2m-maps-50-100} and
\ref{fig:a3m-maps-50-100}.  For the \smica\ and \nilc\ maps their
inpainting masks with $\fsky\approx0.97$ have been used, while for the
\sevem\ map an inpainting mask was not available, so we used its validity mask
with $\fsky=0.76$ instead.  For this reason, the distributions for the
\sevem\ map are significantly broader than the other two.  For the most
part the three maps are in very good agreement though there are notable
exceptions.  In particular $a_{21}^{\rmn{im}}$ and $a_{31}^{\rmn{re}}$ show
significant shifts as compared to their widths between the \smica\ and
\nilc\ maps.  In even more cases the \smica\ and \nilc\ $\alm$ are in the
tails of the \sevem\ distribution.  These results provide a graphical view
of the levels and properties of residual contaminations in the maps.

\subsection{Doppler quadrupole}
\label{ssecn:DQ}

\begin{table}
  \caption{The spherical harmonic coefficients for the quadrupole from the
    \Planck\ inpainted \smica\ map and the Doppler quadrupole (DQ)\@.  The
    $a_{2m}$ values are given in $\muK$. The corrected $a_{2m}$ are
    calculated by subtracting the first two rows.}
  \label{tab:a2m-DQ}
  \begin{tabular}{ld{2}cc} \hline
    Source & \multicolumn{1}{c}{$a_{20}$} & $a_{21}$ & $a_{22}$ \\ \hline
    \Planck\ \smica & 13.09 & $-1.53 + 2.50 \iimag$ & $-15.50 - 17.09
    \iimag$ \\
    DQ  correction & 1.46 & $\hphantom{-}0.28-2.64 \iimag$ &
    $-\hphantom{1}1.16 - \hphantom{1}0.25\iimag$ \\
    \smica\ corrected & 11.63 & $-1.81 + 5.14 \iimag$ & $-14.34 -16.84
    \iimag$ \\
    \hline
  \end{tabular}
\end{table}

The study of the CMB temperature anisotropies typically begins with the
quadrupole and proceeds to higher multipoles, smaller scales.  The monopole
is not included since its magnitude is not predicted by the theory.  The
dipole is subtracted since our motion through the Universe with respect to
the CMB is at a speed $\beta=v/c\sim10^{-3}$, whereas the fluctuations are
$\Delta T/T\sim10^{-5}$ so the Doppler dipole is about two orders of
magnitude larger than the expected cosmological CMB dipole.  A Doppler
quadrupole~(DQ) -- effect of the Sun's proper motion on the quadrupole --
is also induced and expected to have a magnitude
$\mathcal{O}(\beta^2)\sim10^{-6}$.  Though this is small, it is not
negligible, especially in our Universe which has a small cosmological
quadrupole and when properties of the maps are being studied.

The CMB temperature (monopole) has been determined from the nearly perfect
black body to be $T_0=2.7255 \pm 0.0006 \unit{K}$ \citep{Fixsen2009}. Our
direction through the Universe in Galactic coordinates is $(l,b)=(263\fdg
99 \pm 0\fdg 14, 48\fdg 26 \pm 0\fdg 03)$ with a speed $\beta=(1.231\pm
0.003)\sci{-3}$ \citep{wmap5-maps}.  These values can be compared to those
in table~3 of \citet{Planck-R1-V}.  Note that there is an additional
contribution to the speed due to the velocity of the satellite with respect
to the Sun.  This introduces up to about a $10$ per cent variation in
$\beta$ at a non-constant direction with respect to the CMB\@.  This
contribution has not been included in the following discussion.

With these values, the Doppler contribution to the quadrupole from the CMB
monopole may be calculated as \citep{Peebles1968,Kamionkowski2003}
\begin{equation}
  a_{2m}^{\rmn{DQ}} = T_0 \int \left[ \gamma (1-\vec\beta\cdot \unitvec e)
    \right]^{-1} Y_{2 m}^* (\unitvec e) \, \dderiv\unitvec e,
\end{equation}
where $\gamma\equiv(1-|\vec\beta|^2)^{-1/2}$.  The numerical results of this
calculation along with the values from the \smica\ map are given in
Table~\ref{tab:a2m-DQ}.  We see that while the DQ correction is generally
small it is not negligible and in the case of the imaginary part of
$a_{21}$ the correction is comparable to the uncorrected value!  This has
important consequences for any analysis involving the $a_{2m}$, such as the
alignments considered here.

Even higher multipole moments are also induced due to our motion but these
are further suppressed by powers of $\beta$ and need not be considered.
For example, the Doppler octopole should have a magnitude
$\mathcal{O}(\beta^3)\sim10^{-9}$.  Direct calculation shows that for all
components the Doppler octopole correction has a magnitude less than about
$10^{-4}$ of the observed octopole.

In principle the DQ can be corrected during data reduction.  This is
naturally accomplished when calibrating off the anisotropies induced by our
motion with respect to the rest frame of the CMB (typically referred to as
calibrating off the Doppler dipole).  For a blackbody with temperature
$T_0$ and motion in the direction $\unitvec e$ with time-dependent
velocity $\vec\beta(t)$ the induced anisotropies are
\begin{equation}
  \Delta T(\unitvec e,t) = \left[
    \frac{1}{\gamma(t)(1-\vec\beta(t)\cdot\unitvec e)} - 1 \right] T_0.
\end{equation}

For \Planck\ this expression \emph{was} used to calibrate the low-frequency
instrument \citep{Planck-R1-V}.  Surprisingly, for the high-frequency
instrument the classical approximation for the dipole anisotropy
\begin{equation}
  \Delta T(\unitvec e) \approx (\vec\beta\cdot\unitvec e) T_0
\end{equation}
was employed \citep{Planck-R1-VIII}.  \WMAP\ also used this classical
approximation~\citep{WMAP1-data}.  This means that the \WMAP\ ILC maps
require the DQ correction and that it is less clear how to handle the
\Planck\ full-sky maps.  On the one hand, most of the individual frequency
band data combined to produce the cleaned full-sky maps have not been DQ
corrected (the high frequency bands).  On the other hand, one of the
cleanest frequency bands, $70\unit{GHz}$, has been DQ corrected.  Given the
current state, we have applied the DQ correction to all the cleaned
full-sky \Planck\ maps.

Subsequent to our original analysis \citet{Planck-R1-XXIII} was updated to
include estimates of DQ correction factors for each of their released
combined maps.  Unfortunately neither a complete description of how these
correction factors were calculated nor all the data required to calculate
them were made publicly available.  Given the importance of the DQ
correction to alignment results we have recalculated some of the statistics
as discussed in Sec.~\ref{ssecn:results-mpv}.  We find that the \Planck\ DQ
correction factors \emph{strengthen} the alignments making the data
\emph{less consistent} with \LCDM\@.  This further shows the importance of
applying the DQ correction and having an accurate estimate of it.  Due to
the uncertainty in the calculation of the factors we will continue to apply
our simple DQ correction as discussed above and again affirm that the
significance of the anomalous alignments is sensitive to details such as
the DQ correction.

\subsection{Large-scale data summary}

The comparisons presented here provide a broad view of the state of the
existing CMB temperature data on the largest scales.  There are many other
comparisons that can be made (see \citealt{Frejsel2013,Kovacs2013}, for
example).  On the largest scales we see that there exist significant
discrepancies both between the \WMAP\ and \Planck\ maps and even internally
among the maps produced by the \WMAP\ team in different releases and among
the maps produced using the various cleaning methods employed by \Planck.
This suggests that one must be careful about assigning a cosmological
origin to any large-scale result based on full skies.

Despite the caution that must be exercised when interpreting large-scale,
full-sky results, it is important to pursue such work.  The discovery of
alignments, or any anomaly, that persist through multiple full-sky maps is
striking given the different instruments, systematics, cleaning procedures,
etc.  It strongly suggests that there is at least some fundamental origin
to them.  The extent to which this origin is cosmological and the
statistical significance of such an identification is difficult to
determine.  Even if the origin of the alignments is ultimately determined
to be an as-yet unidentified systematic or an unsubtracted (or
mis-subtracted) foreground, it is important to characterize the properties
of the anomalies.  Only in this way can their cause be isolated and (if it
is not cosmological) the data processing pipeline thereby improved,
resulting in higher quality large-scale data.

\section{Alignment statistics}
\label{secn:alignment}

The planarity of the octopole and alignment of the quadrupole and octopole
in the CMB as observed by \WMAP\ was first studied by \citet{dOC-alignment}
through the maximum angular momentum dispersion. This statistic has
subsequently been applied to the \Planck\ data \citep{Planck-R1-XXIII}.  A
more complete picture of CMB alignments is obtained through the use of the
multipole vectors \citet{MPV}.  Here we study both of these approaches.

\subsection{Maximum angular momentum dispersion}

The angular momentum dispersion about an axis $\unitvec n$ is defined by
\citet{dOC-alignment} as
\begin{equation}
  \left[ \Delta L(\unitvec n) \right]_\ell^2 \equiv \sum_m m^2
  |\alm(\unitvec n)|^2.
\end{equation}
Heuristically, $\left[ \Delta L(\unitvec n) \right]_\ell^2 $ measures the
amount of planarity of structure in the multipole $\ell$ around an axis
$\unitvec n$.  Notice that this is not a rotationally invariant quantity
and its value is maximized around some axis, $\unitvec n_\ell$.  Having
found such axes for the quadrupole and octopole we quantify their alignment
through their dot product, $|\unitvec n_2\cdot \unitvec n_3 |$.

To determine the axes $\unitvec n_\ell$ without applying brute force
numerical rotations of maps at some resolution, we follow
\citet{CHSS-anomalies} and use the known rotational properties of the
$\alm$.  Here we briefly review the procedure.  Under rotations the
spherical harmonic coefficients transform as $\vec a'_\ell = \mat D^\dagger
\vec a_\ell$. It is most convenient to represent the rotation in terms of
the Wigner rotation matrices which, for Euler angles $\alpha$, $\beta$, and
$\gamma$, may be written as \citep{Edmonds1960}
\begin{equation}
  D_{m' m}^{(\ell)} (\alpha\, \beta\, \gamma) = \eexp^{\iimag m'\gamma}
  d_{m' m}^{(\ell)}(\beta) \eexp^{\iimag m\alpha}.
\end{equation}
Here the $d_{m' m}^{(\ell)}(\beta)$ are the reduced Wigner matrices. From
explicit calculation we find that
\begin{eqnarray}
  (\Delta L)_\ell^2
  & = & \sum_{m' m''} a_{\ell m'}^* a_{\ell m''} \eexp^{\iimag
    (m'-m'')\gamma}
  \nonumber \\
  & & {} \qquad \times \sum_m m^2 d_{m' m}^{(\ell)}(\beta) d_{m''
    m}^{(\ell)}(\beta) \nonumber \\
  & \equiv & \sum_{m' m''} H_{m' m''}^{(\ell)}(\gamma) G_{m''
    m'}^{(\ell)}(\beta) \\
  & = & \trace\left[ \mat H^{(\ell)}(\gamma) \mat G^{(\ell)}(\beta)
    \right]. \nonumber
\end{eqnarray}
To find the extrema of $(\Delta L)_\ell^2$ we take derivatives with respect
to the angles reducing the problem to the solution of two coupled
non-linear equations
\begin{equation}
  \trace\left[ \upartial_\gamma \mat H^{(\ell)}(\gamma) \mat
    G^{(\ell)}(\beta) \right] = 0, \quad
  \trace\left[ \mat H^{(\ell)}(\gamma) \upartial_\beta \mat
    G^{(\ell)}(\beta) \right] = 0.
\end{equation}
It is easy to see that
\begin{equation}
  \upartial_\gamma H_{m' m''}^{(\ell)}(\gamma) = \iimag (m'-m'')  H_{m'
    m''}^{(\ell)}(\gamma).
\end{equation}
Both $G_{m'' m'}^{(\ell)}(\beta)$ and $\upartial_\beta G_{m''
m'}^{(\ell)}(\beta)$ may be evaluated quickly and efficiently using known
properties of the $d_{m' m}^{(\ell)}(\beta)$ \citep{Edmonds1960}.

We note that for the quadrupole the axis $\unitvec n_2$ may instead be
calculated directly from the multipole vectors as $\unitvec n_2 = \unitvec
w^{(2;1,2)}$ for $\vec w^{(2;1,2)}$ defined below,
Eq.~(\ref{eq:mpv-area}). (For the derivation of this correspondence see
\citealt{CHSS-anomalies}.)

\subsection{Multipole vectors}

The multipole vectors form an irreducible representation of the rotation
group \group{SO(3)} so provide a basis for expanding any scalar function on
the sphere.  These vectors contain all the information about the function
though, in a form different from the usual spherical harmonic coefficients,
$\alm$.  Various properties of the scalar function are more easily
described or detected in one basis than the other, and for considering the
alignments the multipole vectors have much to offer.

Consider a scalar function, $f_\ell(\unitvec e)$, of pure multipole $\ell$.
In the usual spherical harmonic decomposition this function is expanded as
\begin{equation}
  f_\ell(\unitvec e) = \sum_m \alm Y_{\ell m}(\unitvec e).
\end{equation}
In the multipole vector representation it is instead written in terms of a
scalar, $A^{(\ell)}$, and $\ell$ unit vectors, $\left\{ \unitvec
v^{(\ell;j)} \mid j=1,\ldots, \ell \right\}$, as
\begin{equation}
  f_\ell(\unitvec e) = A^{(\ell)} \left[ \prod_{j=1}^\ell \left( \unitvec
    v^{(\ell;j)} \cdot \unitvec e\right) - \mathcal{T}_\ell \right].
\end{equation}
Here $\mathcal{T}_\ell$ is the sum over all possible traces; it removes
the lower order multipoles from the preceding product.  Notice that when
written in this form the function is manifestly rotationally invariant.
Also notice that these vectors are only determined up to a sign so that they
really define axes.

This representation is not new: James Clerk Maxwell in his study of
electromagnetism discussed the properties of spherical harmonics in a very
similar manner \citep{Maxwell-EM}.  It has been shown that Maxwell's
representation is identical to the one given above \citep{Weeks_Maxwell}
suggesting the alternative name of Maxwell's multipole vectors.  With their
introduction into CMB studies an algorithm for converting from the usual
spherical harmonic coefficients, $\alm$, to the multipole vectors was
provided \citep{MPV} greatly facilitating their computation.

There are many ways in which the multipole vectors can be employed to study
the CMB\@.  The most useful way has proven to be through the area vectors
\citep{MPV,CHSS-anomalies}
\begin{equation}
  \vec w^{(\ell;i,j)} \equiv \unitvec v^{(\ell;i)} \times \unitvec
  v^{(\ell;j)}.
  \label{eq:mpv-area}
\end{equation}
These area vectors define sets of planes. Alignment of these planes with a
direction $\unitvec n$ is checked by again using the dot product to define
a set of values $\left\{ A_j \mid j=1,\ldots, n \right\}$ by
\begin{equation}
  A_j \equiv |\vec w_j \cdot \unitvec n|.
\end{equation}
Here $\vec w_j$ represents any area vector.  Different sets of area vectors
can be used in different cases.  For the work discussed here we restrict
the analysis to the quadrupole, $\ell=2$, and octopole, $\ell=3$, vectors.
Finally we define two statistics based on these values
\begin{equation}
  S \equiv \frac1n \sum_{j=1}^n A_j, \quad
  T \equiv 1- \frac1n \sum_{j=1}^n \left( 1 - A_j\right)^2.
  \label{eq:ST-statistics}
\end{equation}
See \citet{CHSS-anomalies} for more general versions of these statistics
and \citet{Weeks_Maxwell, SSHC2004, Slosar2004, Katz2004, Land-axis,
  Land-MPV, Bielewicz2005, Abramo2006-alignments, Abramo2006-anomalies,
  Weeks2007, Gruppuso2009-anomalies, Gruppuso2010} for other tests with
multipole vectors.

\section{Results}
\label{secn:results}

The main results in this work are based on DQ-corrected maps as discussed
in Section~\ref{ssecn:DQ}.  Also, for the \Planck\ \smica\ and \nilc\ maps
the results are based on $5\sci{5}$ harmonic inpaintings using the method
discussed in Section~\ref{ssecn:harmonic_inpainting}.

\subsection{Maximum angular momentum dispersion}

\begin{table}
  \caption{Maximum angular momentum dispersion direction alignments
    reported by \Planck.  The \Planck\ results and $p$-values are from
    table~17 of \protect\citet{Planck-R1-XXIII} and are based on Wiener
    filtered maps.  The uncertainties in these $p$-values are calculated
    based on their use of $1000$ realizations.  The values from this work
    are based on $5\sci{5}$ realizations.  See the text for details.}
  \label{tab:Lz2-Planck}
  \begin{tabular}{ld{1.3}D{,}{\pm}{2.2}d{1.2}} \\ \hline
    & & \multicolumn{2}{c}{$p$-value (per cent)} \\
    Map
    & \multicolumn{1}{c}{$|\unitvec n_2\cdot\unitvec n_3|$}
    & \multicolumn{1}{c}{\Planck}
    & \multicolumn{1}{c}{This work}
    \\ \hline
    \Planck\ \nilc & 0.974 & 3.3,0.6 & 2.59 \\
    \Planck\ \sevem & 0.988 & 1.6,0.4 & 1.17 \\
    \Planck\ \smica & 0.977 & 3.2,0.6 & 2.28 \\
    \hline
  \end{tabular}
  \end{table}

\begin{table}
  \caption{Maximum angular momentum dispersion direction alignments for
    maps with and without DQ correction.  See the text for details.}
  \label{tab:Lz2-maps-DQ}
  \begin{tabular}{ld{1.4}d{1.3}d{1.4}d{1.3}}
    \\ \hline
    & \multicolumn{2}{c}{Uncorrected} & \multicolumn{2}{c}{DQ corrected} 
    \\
    Map
    & \multicolumn{1}{c}{$|\unitvec n_2\cdot\unitvec n_3|$}
    & \multicolumn{1}{c}{$p$-value (\%)}
    & \multicolumn{1}{c}{$|\unitvec n_2\cdot\unitvec n_3|$}
    & \multicolumn{1}{c}{$p$-value (\%)}
    \\ \hline
    \WMAP\ ILC 7yr & 0.9999 & 0.006 & 0.9966 & 0.327
    \\
    \WMAP\ ILC 9yr & 0.9985 & 0.150 & 0.9948 & 0.511 
    \\
    \Planck\ \nilc & 0.9902 & 0.955 & 0.9988 & 0.118
    \\
    \Planck\ \sevem & 0.9915 & 0.825 & 0.9995 & 0.055
    \\
    \Planck\ \smica & 0.9809 & 1.883 & 0.9965 & 0.338
    \\
    \hline
  \end{tabular}
\end{table}

The alignment of the quadrupole and octopole as quantified by the maximum
angular momentum dispersion axes has been studied in recent data releases
\citep{WMAP7-anomalies,WMAP9-results,Planck-R1-XXIII}.
\citet{Planck-R1-XXIII} used Wiener filtered maps to quantify the alignment
through the dot product, $|\unitvec n_2\cdot\unitvec n_3|$, and determined
the fraction of realizations with at least this level of alignment in
\LCDM\ from $1000$ realizations.  The results reported by \Planck\ are
summarized in Table~\ref{tab:Lz2-Planck}; we have roughly estimated the
uncertainty in these $p$-values for the \Planck\ analysis based on the
simplifying assumption of Poisson statistics.  For comparison, using the
$|\unitvec n_2\cdot\unitvec n_3|$ values provided by \Planck\ we have
recalculated the $p$-values based on $5\sci{5}$ realizations of \LCDM\@.
These values are also included in Table~\ref{tab:Lz2-Planck} and are
consistently a little more than one-sigma lower than those provided by
\Planck. While the number of simulations used by the \Planck\ collaboration
is significantly smaller than our sample, their simulations also include
instrumental effects and ours do not. The small difference in $p$-values
suggests that instrumental effects are not dominant and add at most a small
correction of the order of the statistical uncertainty of the
\Planck\ simulation itself.

To study the effect of the extra cleaning provided by the Wiener filtering
we have calculated the maximum angular momentum dispersion axes from the
full-sky maps provided by \Planck.  For the \smica\ and \nilc\ maps the
\Planck\ inpainted maps were analysed.  The results are given in
Table~\ref{tab:Lz2-maps-DQ} as the `Uncorrected' values.  These $p$-values
should be compared to our $p$-values from Table~\ref{tab:Lz2-Planck} (last
column).  We see that the provided maps exhibit somewhat more alignment
than the Wiener filtered maps.  So something \emph{has} been removed by the
Wiener filtering.  Whether it is noise or CMB signal is unclear.

Also included in Table~\ref{tab:Lz2-maps-DQ} are the results from the
\WMAP\ data releases.  The `Uncorrected' values are consistent with
discussions in the \WMAP\ seven-year \citep{WMAP7-anomalies} and nine-year
\citep{WMAP9-results} analyses.  The alignment in the seven-year data is
quite remarkable for being almost perfect ($|\unitvec n_2\cdot\unitvec
n_3|\simeq 1$).  The change in alignment in the nine-year data is largely
attributed to improvements in the asymmetric beam deconvolution
\citep{WMAP9-results} and is one example of how analysis improvements
affect the alignments.  Even so, the \WMAP\ maps show more alignment than
the \Planck\ maps.

The effect of the Doppler quadrupole correction as discussed in
Sec.~\ref{ssecn:DQ} is also included in Table~\ref{tab:Lz2-maps-DQ}.
Interestingly, the DQ correction has the opposite effect on the \WMAP\ and
\Planck\ alignments.  Since the \WMAP\ alignments are so precise this
correction lessens the significance as we would expect, however, for
\Planck\ we find the alignments become more significant.  More importantly,
\WMAP\ and \Planck\ are found to be in \emph{better} agreement with each
other after the DQ correction has been applied.

\begin{figure*}
  \includegraphics[width=7in]{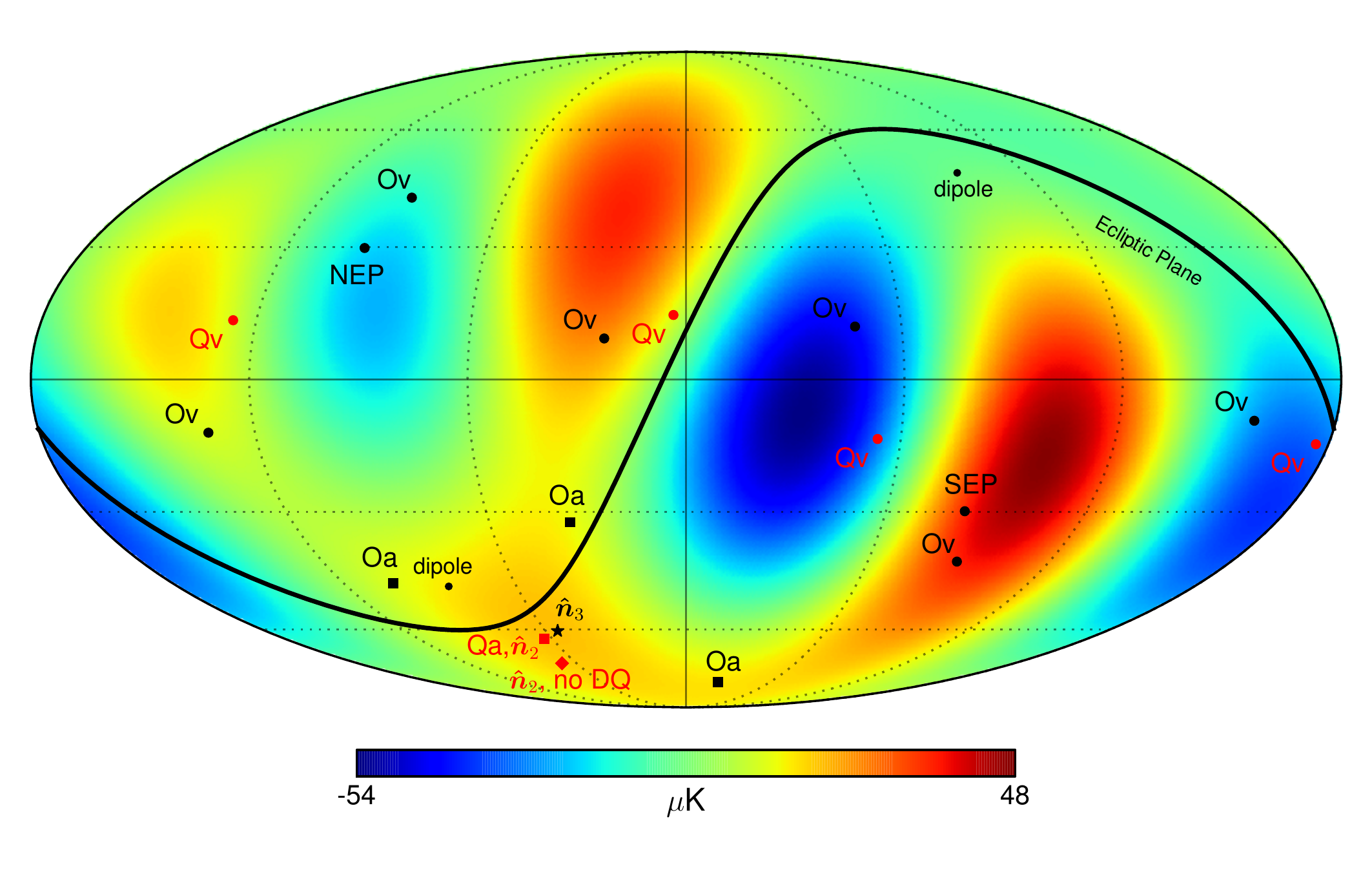}
  \caption{Quadrupole and octopole multipole vectors for the DQ corrected
    \smica\ map in Galactic coordinates. The background shows the
    quadrupole+octopole pattern from the DQ corrected \smica\ map.  The
    multipole vectors are shown as circles, in red and labelled `Qv' for
    the quadrupole and in black and labelled `Ov' for the octopole. The
    direction of the area vectors defined in Eq.~(\ref{eq:mpv-area}),
    $\unitvec w^{(\ell;i,j)}$, are shown as squares.  Again the quadrupole
    area vector is in red and labelled `Qa' and the octopole area vectors
    are in black and labelled `Oa'.  Since the multipole vectors are only
    determined up to a sign each vector appears twice in the figure.  The
    area vectors have only been plotted in the southern hemisphere to avoid
    cluttering the plot.  The maximum angular momentum dispersion direction
    for the octopole, $\unitvec n_3$, is shown as the black star.  Since
    $\unitvec n_2=\unitvec w^{(2;1,2)}$ it is also represented by the red
    square.  The direction of $\unitvec n_2$ without the DQ correction is
    shown as the red diamond. For reference also shown in the figure is the
    Ecliptic plane (black line), the locations of the north (NEP) and south
    (SEP) Ecliptic poles, and the direction of our motion with respect to
    the CMB (dipole). The coordinates of the vectors are listed in
    Table~\ref{tab:smica-mpv}.}
  \label{fig:mpv}
\end{figure*}

\begin{table}
  \caption{Average DQ-corrected multipole vectors from $5\sci{5}$ harmonic
    inpaintings of the \smica\ map.  The vector directions are given in
    Galactic coordinates, $(l,b)$, and their Cartesian equivalents,
    $(x,y,z)$.  These vectors are plotted in Fig.~\ref{fig:mpv}.}
  \label{tab:smica-mpv}
  \begin{tabular}{ld{3.1}d{3.1}d{1.3}d{1.3}d{1.3}c}
    \\ \hline
    Vector
    & \multicolumn{1}{c}{$l$ (deg)}
    & \multicolumn{1}{c}{$b$ (deg)}
    & \multicolumn{1}{c}{$x$}
    & \multicolumn{1}{c}{$y$}
    & \multicolumn{1}{c}{$z$}
    & Magnitude
    \\ \hline
    $\unitvec v^{(2,1)}$ & 3.5 & 14.4 & 0.967 & 0.059 & 0.249 & -- \\
    $\unitvec v^{(2,2)}$ & 126.5 & 13.3 & -0.579 & 0.783 & 0.229 & -- \\
    $\vec w^{(2;1,2)}$   & 63.6 & -62.7 & 0.182 & 0.366 & -0.791 & 0.890 \\
    $\unitvec v^{(3,1)}$ & 90.5 & 42.0 & -0.007 & 0.744 & 0.669 & -- \\
    $\unitvec v^{(3,2)}$ & 22.6 & 9.2 & 0.911 & 0.380 & 0.159 & -- \\
    $\unitvec v^{(3,3)}$ & -47.1 & 11.8 & 0.667 & -0.716 & 0.205 & -- \\
    $\vec w^{(3;1,2)}$  & 102.5 & -47.4 & -0.136 & 0.610 & -0.680 & 0.924 \\
    $\vec w^{(3;2,3)}$  & -22.8 & -77.0 & 0.192 & -0.081 & -0.906 & 0.930 \\
    $\vec w^{(3;3,1)}$  & 35.3 & -32.4 & 0.632 & 0.447 & -0.491 & 0.916 \\
    \hline
  \end{tabular}
\end{table}

\subsection{Multipole vectors}
\label{ssecn:results-mpv}

\begin{table*}
  \caption{The $S$ and $T$ alignment statistics from
    Eq.~(\ref{eq:ST-statistics}) for various directions.  Listed are the
    $p$-values in per cent of \LCDM\ producing a value larger than that
    found in the given map based on $10^6$ realizations of \LCDM.  The
    directions tested are the quadrupole+octopole alignment (Q+O), the
    Ecliptic plane, the north Galactic pole (NGP), and the direction of our
    motion with respect to the CMB (dipole).  The results for the
    \smica\ and \nilc\ maps are based on the average of the $S$ statistic
    from $5\sci{5}$ harmonic inpaintings of these maps. The results below
    the '\Planck\ DQ Correction' line were calculated using the \Planck\ DQ
  correction factors as discussed in the text.}
  \label{tab:mpv-ST}
  \begin{tabular}{ld{2}d{2}d{2}d{2}d{2}d{2}d{2}d{2}}
    \\ \hline
    & \multicolumn{2}{c}{Q+O} & \multicolumn{2}{c}{Ecliptic Plane} &
    \multicolumn{2}{c}{NGP} & \multicolumn{2}{c}{dipole} \\
    Map
    & \multicolumn{1}{c}{$S$}
    & \multicolumn{1}{c}{$T$}
    & \multicolumn{1}{c}{$S$}
    & \multicolumn{1}{c}{$T$}
    & \multicolumn{1}{c}{$S$}
    & \multicolumn{1}{c}{$T$}
    & \multicolumn{1}{c}{$S$}
    & \multicolumn{1}{c}{$T$}
    \\ \hline
    \WMAP\ ILC 7yr  & 0.22 & 0.10 & 2.66 & 2.70 & 0.82 & 0.90
    & 0.18 & 0.20 \\
    \WMAP\ ILC 9yr  & 0.18 & 0.08 & 1.96 & 1.82 & 0.79 & 0.76
    & 0.14 & 0.15 \\
    \Planck\ \nilc  & 1.85 & 1.05 & 2.80 & 3.04 & 1.41 & 1.26
    & 0.32 & 0.19 \\
    \Planck\ \sevem & 0.41 & 0.22 & 2.52 & 2.94 & 0.79 & 0.92
    & 0.09 & 0.05 \\
    \Planck\ \smica & 1.62 & 0.93 & 3.74 & 4.16 & 1.56 & 1.52
    & 0.37 & 0.30 \\
    \hline
    \multicolumn{9}{c}{\Planck\ DQ Correction} \\
    \hline
    \Planck\ \nilc  & 0.54 & 0.27 & 2.55 & 2.64 & 1.14 & 1.10
    & 0.18 & 0.14 \\
    \Planck\ \sevem & 0.16 & 0.08 & 2.31 & 2.58 & 0.73 & 0.89
    & 0.06 & 0.05 \\
    \Planck\ \smica & 0.54 & 0.28 & 3.37 & 3.60 & 1.32 & 1.39
    & 0.23 & 0.23 \\
    \hline
  \end{tabular}
\end{table*}

\begin{figure}
  \includegraphics[width=3.5in]{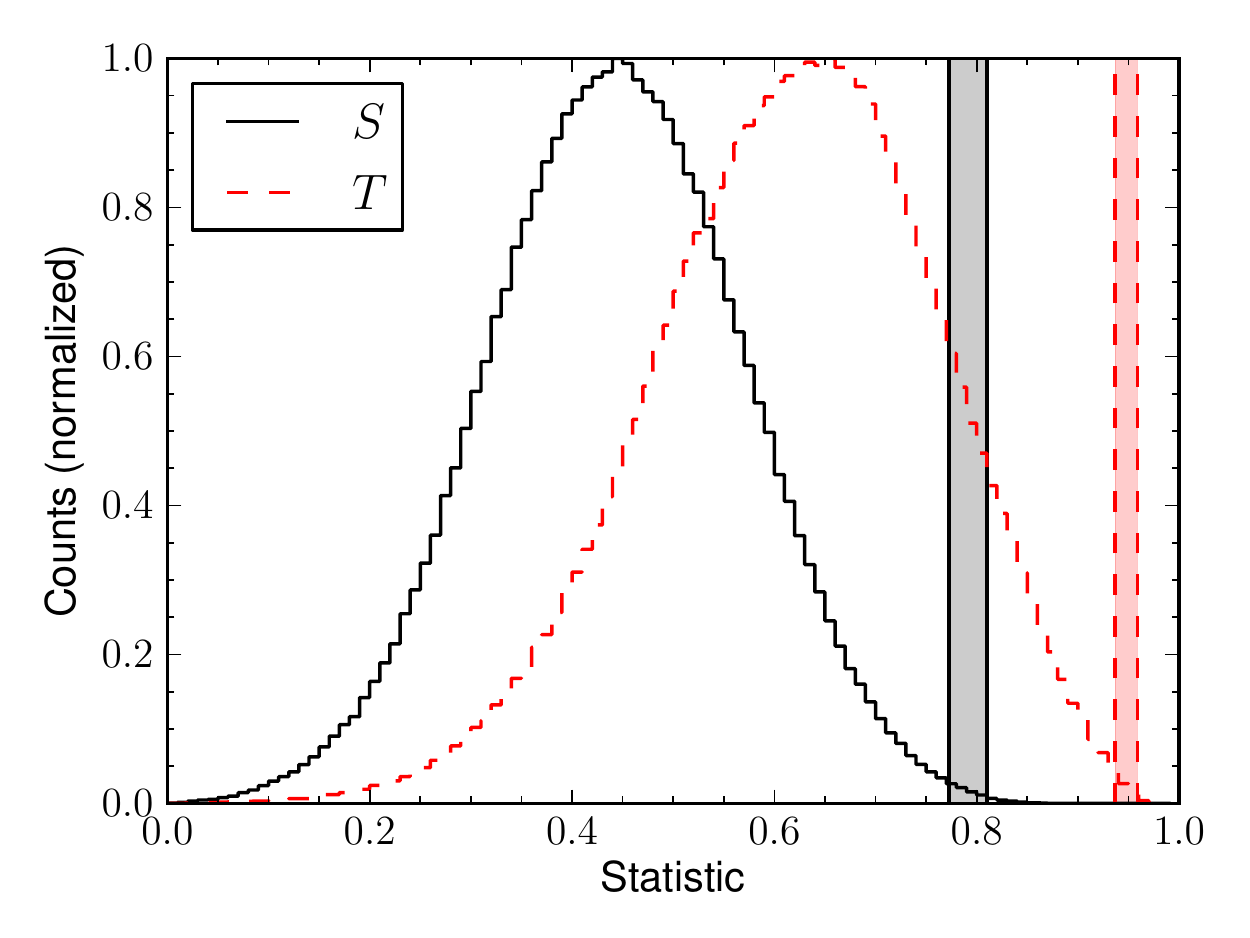}
  \caption{The $S$ and $T$ statistics from Eq.~(\ref{eq:ST-statistics}) for
    the alignment of the multipole vectors with the direction of our motion
    with respect to the CMB (the dipole direction). The histograms
    represent the distribution of the $S$ (solid, black line) and $T$
    (dashed, red line) statistics from $10^6$ realizations of \LCDM\@.
    The shaded regions between the vertical lines represent the range of
    values found for the CMB maps studied in this work.  See
    Table~\ref{tab:mpv-ST} for the full results.}
  \label{fig:mpv-dipole}
\end{figure}

The multipole vectors for the DQ corrected \smica\ map are listed in
Table~\ref{tab:smica-mpv} and shown in Fig.~\ref{fig:mpv} plotted in
Galactic coordinates.  When compared to fig.~3 from \citet{CHSS-anomalies},
this figure shows that the general features have not changed significantly
since the first-year \WMAP\ data release.  It also provides a visual
summary of many of the large-angle anomalies.  In particular,
\begin{enumerate}
\item the Ecliptic plane is seen to carefully thread itself between a hot
  and cold spot and there is a clear power asymmetry across the Ecliptic
  plane;
\item the planarity of the octopole and the alignment of the quadrupole and
  octopole planes is clearly visible -- note the remarkable near-overlap of
  the quadrupole and octopole maximum angular momentum dispersion axes;
\item the area vectors lie near each other, near the Ecliptic plane, and
  also near the dipole direction.
\end{enumerate}

To quantify the alignments we consider the mutual alignment of the
quadrupole and octopole area vectors as well as alignments with the
Ecliptic plane, north Galactic pole (NGP), and the direction of our motion
with respect to the CMB (dipole).  This is a subset of the directions
considered in \citet{CHSS-anomalies}.  The results based on the $S$ and $T$
statistics from Eq.~(\ref{eq:ST-statistics}) are shown in
Table~\ref{tab:mpv-ST}, and indicate that alignments persist at the $95$ to
$99.9$ per cent level, with the strongest alignment occurring with the
dipole direction ($\geq 99.6$ per cent). As we argued above, the spread of
values between the different maps gives an idea of the effect of the
residual systematic errors due to foregrounds.

As noted above, subsequent to our original analysis \citet{Planck-R1-XXIII}
was updated to include estimates for DQ correction factors based on the
calibration techniques and weights applied to the individual frequency band
maps combined to create the cleaned, full-sky maps.  The correction factors
were found to be $1.7$, $1.5$, and $1.7$ for the \Planck\ \nilc, \sevem,
and \smica\ maps, respectively.  To determine the effect of using these
correction factors we have reanalysed the alignments of the multipole
vectors for these three maps.  For the \Planck\ \smica\ map the direction
of the oriented area vector for the quadrupole based on our simple DQ
correction ($\unitvec n_2$ and labelled `Qa' in Fig.~\ref{fig:mpv}) moves
about $3.2$ degrees when calculated using the \Planck\ DQ correction.  This
is a significantly smaller change than that from the vector without any DQ
correction applied (the `no DQ' vector in Fig.~\ref{fig:mpv}.  Of more
importance, the magnitude of the oriented area vector, $\vec w^{(2;1,2)}$,
has changed from $0.890$ (see Table~\ref{tab:smica-mpv}) to $0.948$.  In
other words, with the \Planck\ DQ correction of the \Planck\ \smica\ map
the quadrupole multipole vectors are even more nearly perpendicular.  The
effect this has on the $S$ and $T$ statistics for the alignments we have
studied is given in Table~\ref{tab:mpv-ST} under the `\Planck\ DQ
Correction' line.  In all cases we see that the \Planck\ DQ correction
leads to alignments \emph{more unlikely} in \LCDM\@.

To make the anomalous nature of the alignments more clear the results for
the $S$ and $T$ statistics from Eq.~(\ref{eq:ST-statistics}) of the
multipole vectors with the direction of our motion with respect to the CMB
(the dipole direction) are shown in Fig.~\ref{fig:mpv-dipole}.  In this
figure the histograms are the expected values from \LCDM\ and the vertical
lines the range of values found for the various maps.  In all cases the
observed alignments reside far in the tail of the expected distributions.

\begin{table}
  \caption{The alignment $p$-value based on the $S$ and $T$
    statistics from Eq.~(\ref{eq:ST-statistics}) \emph{given the observed
    quadrupole+octopole alignment}.  The values are quoted in per cent and
    are based on $10^6$ randomly rotated directions. See the text for
    details.}
  \label{tab:area-random-directions}
  \begin{tabular}{ld{2}d{2}d{2}d{2}d{2}d{2}}
    \\ \hline
    & \multicolumn{2}{c}{Ecliptic Plane} &
    \multicolumn{2}{c}{NGP} & \multicolumn{2}{c}{dipole} \\
    Map
    & \multicolumn{1}{c}{$S$}
    & \multicolumn{1}{c}{$T$}
    & \multicolumn{1}{c}{$S$}
    & \multicolumn{1}{c}{$T$}
    & \multicolumn{1}{c}{$S$}
    & \multicolumn{1}{c}{$T$}
    \\ \hline
    \WMAP\ ILC 7yr  & 3.27 & 3.33 & 12.26 & 12.74 & 6.47 & 6.55
    \\
    \WMAP\ ILC 9yr  & 1.30 & 1.38 & 12.56 & 12.35 & 6.02 & 6.30
    \\
    \Planck\ \nilc  & 4.39 & 5.14 & 12.26 & 12.74 & 5.50 & 5.08
    \\
    \Planck\ \sevem & 4.93 & 6.37 & 11.75 & 12.74 & 3.98 & 3.17
    \\
    \Planck\ \smica & 5.17 & 6.10 & 12.66 & 13.15 & 5.94 & 6.05
    \\
    \hline
  \end{tabular}
\end{table}

\subsection{Interdependence of alignments}

The alignments shown in Table~\ref{tab:mpv-ST} are peculiar, particularly
the mutual alignment of the quadrupole and octopole area vectors as well as
that with the dipole direction.  A remaining question is the independence
of these alignments.  

As a first method of addressing this, we follow
\citet{CHSS-anomalies} and calculate the significance of alignments with
various directions \emph{given} the observed relative orientation of the
quadrupole and octopole area vectors.  That is, given the observed
quadrupole and octopole structure, the overall orientation of this pattern
on the sky is arbitrary.  With this in mind we can address how likely it is
for a random orientation of this structure to have the quadrupole and
octopole area vectors at least as well aligned with a fixed direction in
the sky as observed.  Though we think of this as rotating the map, it is
equivalent to comparing the alignments to random directions.  The result of
such a comparison based on $10^6$ random directions using the $S$
and $T$ statistics~(\ref{eq:ST-statistics}) is given in
Table~\ref{tab:area-random-directions}. We find that the residual $p$-value
for alignment with the dipole and Ecliptic Plane directions, given the
mutual quadrupole+octopole alignment, is at the $2$ to $6$ per cent level,
while the alignment with the Galactic pole is not significant. These
results are in agreement with results in \citet{CHSS-anomalies} (see table 4
in that paper), and indicate that even given the relative location of the
quadrupole and octopole area vectors (i.e.\ their mutual alignment), the
Ecliptic plane and dipole alignments are unlikely at the $95$ per cent
level.

\begin{table}
  \caption{The alignment $p$-values based on the $S$ and $T$ statistics
    from Eq.~(\ref{eq:ST-statistics}) \emph{given at least the observed
      quadrupole+octopole alignment}.  The values are quoted in per cent
    and based on $10^6$ realizations of \LCDM\ with at least the
    quadrupole+octopole alignment observed in the \Planck\ \nilc\ map. See
    the text for details.}
  \label{tab:ST-statistics-conditional-q+o}
  \begin{tabular}{ld{1}d{1}d{1}d{1}d{1}d{1}}
    \\ \hline
    & \multicolumn{2}{c}{Ecliptic Plane} &
    \multicolumn{2}{c}{NGP} & \multicolumn{2}{c}{dipole} \\
    Map
    & \multicolumn{1}{c}{$S$}
    & \multicolumn{1}{c}{$T$}
    & \multicolumn{1}{c}{$S$}
    & \multicolumn{1}{c}{$T$}
    & \multicolumn{1}{c}{$S$}
    & \multicolumn{1}{c}{$T$}
    \\ \hline
    \WMAP\ ILC 7yr  & 20.9 & 18.9 & 14.3 & 17.4 & 8.7 & 10.8
    \\
    \WMAP\ ILC 9yr  & 18.6 & 16.5 & 14.3 & 16.8 & 7.9 & 10.0
    \\
    \Planck\ \nilc  & 14.0 & 14.0 & 13.6 & 14.9 & 6.9 & 6.2
    \\
    \Planck\ \sevem & 18.8 & 18.2 & 13.5 & 16.6 & 5.9 & 5.5
    \\
    \Planck\ \smica & 17.4 & 17.1 & 14.4 & 16.3 & 7.8 & 8.1
    \\
    \hline
  \end{tabular}
\end{table}

A related question is to ask in realizations of \LCDM\ with \emph{at least}
the observed alignment of the quadrupole and octopole, what is the
$p$-value for alignment with other directions?  The difference between this
and the previous question is that the orientation of the quadrupole and
octopole is no longer a rigid structure.  From Table~\ref{tab:mpv-ST} we
see that realizations of \LCDM\ with the quadrupole and octopole aligned as
closely as in the data are rare.  The most frequent alignment occurs for
the \Planck\ \nilc\ map with $p$-values of $1.85$ and $1.05$ per cent for
the $S$ and $T$ statistics, respectively.  Using these values as cutoffs,
we generate $10^6$ realizations of \LCDM\ with at least this much
alignment, that is, realizations with either a $S$ or $T$
quadrupole+octopole statistic at least the value from the
\Planck\ \nilc\ map.  Note that this means that the number of realizations
vary for each map.  The ratio of the number of realizations between each
pair of maps is given by the ratio of $p$-values from the Q+O columns in
Table~\ref{tab:mpv-ST}.  For each map we then find the conditional
probability of alignment with some direction, such as the dipole,
\emph{given} at least the observed alignment of the quadrupole+octopole.
The resulting $p$-values are given in
Table~\ref{tab:ST-statistics-conditional-q+o}.  In comparison with
Table~\ref{tab:area-random-directions} we see by relaxing the rigid
structure of the quadrupole and octopole orientations the $p$-values have
increased.  Overall the $p$-values are now roughly $5$ to $21$ per cent
with the more unlikely alignment being with the dipole direction, meaning
that the quadrupole+octopole and dipole alignments are the least
correlated.

The results from Tables~\ref{tab:mpv-ST} and
\ref{tab:ST-statistics-conditional-q+o} can be combined using Bayes'
theorem to find the conditional probability for any two alignments.  Let
$d_1$ and $d_2$ represent two different alignments, then, $P(d_1|d_2)$ is
the conditional probability of alignment with $d_1$ given the alignment with
$d_2$ and similarly for $P(d_2|d_1)$.  These alignments are related as
\begin{equation}
  P(d_1|d_2)P(d_2) = P(d_2|d_1)P(d_1).
\end{equation}
From the tables provided we may immediately calculate the conditional
probability for the alignment of the quadrupole+octopole given the
alignment of any other direction.  To do so let $d_2\equiv\rmn{qo}$
represent the quadrupole+octopole alignment, then
\begin{equation}
  P(\rmn{qo}|d_1) = P(d_1|\rmn{qo})\frac{P(\rmn{qo})}{P(d_1)}.
\end{equation}
Here both $P(\rmn{qo})$ and $P(d_1)$ may be read from
Table~\ref{tab:mpv-ST} and $P(d_1|\rmn{qo})$ from
Table~\ref{tab:ST-statistics-conditional-q+o}.  These conditional
probabilities are shown in Table
\ref{tab:ST-statistics-conditional-direction}. Conditional probabilities
for other pairs of alignments can be calculated from repeated application
of Bayes' theorem.

\begin{table}
  \caption{The quadrupole+octopole alignment $p$-values based on the $S$
    and $T$ statistics from Eq.~(\ref{eq:ST-statistics}) \emph{given at
      least the observed dipole, Ecliptic or Galactic alignment}.  The
    values are quoted in per cent and are obtained using Bayes' theorem and
    the data from Tables \ref{tab:mpv-ST} and
    \ref{tab:ST-statistics-conditional-q+o}.}
  \label{tab:ST-statistics-conditional-direction}
  \begin{tabular}{ld{1}d{1}d{1}d{1}d{1}d{1}}
    \\ \hline
    & \multicolumn{2}{c}{Ecliptic Plane} &  \multicolumn{2}{c}{NGP} &
    \multicolumn{2}{c}{dipole} \\
    Map
    & \multicolumn{1}{c}{$S$}
    & \multicolumn{1}{c}{$T$}
    & \multicolumn{1}{c}{$S$}
    & \multicolumn{1}{c}{$T$}
    & \multicolumn{1}{c}{$S$}
    & \multicolumn{1}{c}{$T$}
    \\ \hline
    \WMAP\ ILC 7yr  & 1.7 & 0.7 & 3.8 & 1.9 & 10.6 & 5.4
    \\
    \WMAP\ ILC 9yr  & 1.7 & 0.7 & 3.3 & 1.8 & 10.2 & 5.3
    \\
    \Planck\ \nilc  & 9.3 & 4.8 & 17.8 & 12.4 & 39.9  & 34.3 
     \\
    \Planck\ \sevem & 3.1 & 1.4 & 7.0 & 4.0 & 26.9 & 24.2
    \\
    \Planck\ \smica  & 7.5 & 3.8 & 15.0 & 10.0 & 34.2 & 25.1
    \\
    \hline
  \end{tabular}
\end{table}

If we assume the observed levels of alignment with the Ecliptic plane or
the Galactic poles, the quadrupole+octopole alignment remains anomalous at
the $1$ to $4$ per cent level for \WMAP. For \Planck\ the conclusion is not
clear, however all $p$-values remain below $20$ per cent.  The situation is
different for the dipole. We see that assuming the observed level of dipole
alignment, the quadrupole+octopole alignment seems to be quite
plausible. This suggests that the dipole alignment (which is also the most
significant and robust alignment in Table \ref{tab:mpv-ST}) could be the
reason for the other observed alignments.

\section{Conclusions}
\label{secn:Conclusions}

The largest structures in the microwave sky, the quadrupole and octopole,
are aligned with one another and with physical directions or planes -- the
dipole direction and the Ecliptic plane.  These alignments, first observed
and discussed in the one-year \WMAP\ data, have persisted throughout
\WMAP's subsequent data releases, and are now confirmed in the one-year
\Planck\ data.  On the one hand, this is to be expected: the largest scales
are precisely measured and the same CMB sky is observed by both satellites.
On the other hand, this is surprising: cleaned, full-sky maps are required
to see these alignments, and the removal of foregrounds, along with other
systematic effects, makes it challenging to accurately produce full-sky
maps on large angular scales.

In this work we have studied the ILC maps from the seven and nine-year
\WMAP\ data releases and the \nilc, \sevem, and \smica\ cleaned maps from
the first-year \Planck\ data release.  Qualitatively, the main anomalies
detected in earlier \WMAP\ releases remain: the quadrupole and octopole are
aligned with each other; the normal to their average plane is aligned with
the dipole -- the direction of our motion through the Universe; that normal
is also close to the Ecliptic plane, so that the average plane of the
quadrupole and octopole is nearly perpendicular to the Ecliptic plane.
Finally, as can be seen in Fig.~\ref{fig:mpv} which shows the sum of
the quadrupole and octopole from the \smica\ map, the Ecliptic plane
cleanly cuts between a hot and cold spot, thereby separating weaker
quadrupole+octopole power in the north Ecliptic hemisphere from the
stronger power in the south Ecliptic hemisphere.

Quantitatively, statistics from the maximum angular momentum dispersion
(Table~\ref{tab:Lz2-maps-DQ}) and the multipole vectors
(Table~\ref{tab:mpv-ST} and Fig.~\ref{fig:mpv-dipole}) both show strong
evidence for the mutual alignment of the quadrupole and the octopole. The
$p$-values for at least as much alignment as observed occurring in
realizations of \LCDM\ are typically less than $0.5$ per cent, once one has
Doppler-corrected the quadrupole.  The exceptions are the \nilc\ and
\smica\ multipole vector statistics, $S$ and $T$, which have $p$-values of
$1$ to $2$ per cent.  These results are strengthened when the \Planck\ DQ
corrections are applied instead of our simple one.

The alignment of the quadrupole and octopole with the dipole
(Table~\ref{tab:mpv-ST}) appears at first sight even more robust than their
mutual alignments, with $p$-values of less than $0.4$ per cent (and as low
as $0.05$ per cent) in all maps and with both $S$ and $T$ statistics.  The
interpretation however is not clear.  The dipole includes contributions
from several sources, but is almost certainly dominated by the Doppler
effect from the Sun's motion through the Galaxy, the Galaxy's motion
through the Local Group, and the Local Group's motion through the more
distant large scale structure; all giving comparable contributions.  (A
dominant or even significant contribution from a cosmological dipole seems
remote.)  Since these contributions originate from gravity gradients on
very different scales, it is difficult to envision physics that would
connect the dipole, quadrupole, and octopole.  A systematic error in the
measurement or the analysis pipeline could connect them all, but the
robustness of the alignment across the two satellites argues against that
explanation.  (A remote possibility does still remain since
\Planck\ calibrated off the \WMAP\ dipole in their first-year data
release.)

As an attempt to disentangle correlation from causation between and among
the alignments, we have studied their interdependence by calculating the
conditional probability of alignment with a fixed direction given the
observed mutual quadrupole+octopole alignment and \emph{vice versa}.
Through application of Bayes' theorem, the conditional probability for any
two alignments can be deduced from these calculations.  For example, the
conditional $p$-values for quadrupole+octopole+dipole alignment given
either the observed quadrupole+octopole alignment (Table
\ref{tab:area-random-directions}) or at least the observed
quadrupole+octopole alignment (Table
\ref{tab:ST-statistics-conditional-q+o}) are $3$ to $10$ per cent.  These
are consistent with the $4$ to $6$ per cent found in \citet{CHSS-anomalies}
for the conditional $p$-value of the quadrupole+octopole+dipole alignment
given the observed relative directions of the quadrupole and octopole area
vectors.  In other words, the alignments with the dipole may well be a
distraction -- a statistical accident -- barring an unknown common
\WMAP-\Planck\ systematic which somehow causes the
dipole+quadrupole+octopole alignment.

\textit{A priori}, less scepticism could be attached to a possible physical
explanation for the correlation between the quadrupole+octopole and the
Ecliptic plane.  If the underlying cosmological quadrupole and octopole
were unexpectedly absent, then we could well imagine a Solar System (or
even nearby Solar-neighbourhood) source for the quadrupole and octopole
correlated with the plane of the Solar System. Nevertheless, there are no
proposed viable physical models that correctly reproduce the observed
arrangement of quadrupole and octopole extrema lying on a plane
perpendicular to the Ecliptic and well separated by it.  The statistics
offer only weak, and even confusing, guidance as to what is correlation and
what, if anything, is causation.  The statistical situation is no clearer
for correlation with the Galactic pole.

Unfortunately, the fact that the dipole direction simply happens to be just
off the Ecliptic plane, which passes about $30$ degrees from the Galactic
poles, makes establishing the priority of one correlation over another
difficult just on the basis of statistics of CMB temperature data.  Some,
or all, of these correlations are presumably accidental. Solving this
puzzle will require data other than just CMB temperature maps, and probably
a model that can be tested against such data.

The lack of correlation on large angular scales from cut-sky maps has been
presented in a companion work \citep{CHSS-Planck-R1-ctheta}.  A natural
next step would be to also study the interdependence of the alignments of
the full-sky low multipoles and the lack of correlations. For the
\WMAP\ releases up to year seven, these were discussed in \citet{Rakic2007}
and \citet{SHCSS2011}.  The corresponding analysis for the final \WMAP\ and
first-year \Planck\ releases will be presented elsewhere, but we expect
that the conclusion from \citet{SHCSS2011}, namely that the lack of angular
correlation and the alignments are uncorrelated anomalies, will remain
valid.

In summary, the quadrupole and octopole alignments noted in early
\WMAP\ full-sky maps persist in the \WMAP\ seven-year and final (nine-year)
maps, and in the \Planck\ first-year full-sky maps.  The correlation of the
quadrupole and octopole with one another, and their correlations with other
physical directions or planes -- the dipole, the Ecliptic, the Galaxy --
remain broadly unchanged across all of these maps.  Consequently, it is not
sufficient to argue that they are less significant than they appear merely
by appealing to the uncertainties in the full-sky maps -- such
uncertainties are presumably captured in the range of foreground removal
schemes that went into the map making. It similarly seems contrived that
the primordial CMB at the last scattering surface is correlated with the
local structures imprinted via the Integrated Sachs-Wolfe (ISW) effect in
just such a way to generate the observed alignments, as proposed elsewhere
\citep{Rakic2006a,Francis2010,Dupe2011,Rassat_Starck_Dupe,Rassat2013}, even
taking for granted the reliability of the procedure to subtract the ISW
signal from the map.

While it may be tempting to explain away the observed large-angle
alignments in the CMB by postulating additional, unspecified corrections to
the maps, such explanations so far have not been compelling.  Numerous
corrections have been applied in the data analysis pipelines, and they have
also evolved between the initial \WMAP\ data releases and the
\Planck\ first-year release, yet the alignments remain. Further, almost
anything done to the maps will lessen the significance of the observed
alignments, so, just because a new correction \emph{could} affect the
observed alignments, this does not mean that an otherwise unspecified new
correction \emph{must} exist.  We think it is preferable to acknowledge
that the existence of anomalies seen in the \WMAP\ and \Planck\ maps at
large angular scales \emph{may} point to residual contamination in the data
or to interesting new fundamental physics.

\section*{Acknowledgements}

We acknowledge valuable communications and discussions with F.~Bouchet,
C.~Burigana, J.~Dunkley, G.~Efstathiou, K.~Ganga, P.~Naselsky, H.~Peiris, 
C.~R\"ath and D.~Scott. 
GDS and CJC are supported by a grant from the US Department of Energy to
the Particle Astrophysics Theory Group at CWRU\@.  DH has been supported by
the DOE, NSF, and the Michigan Center for Theoretical Physics.  DJS is
supported by the DFG grant RTG 1620 `Models of gravity'. DH thanks the
Kavli Institute for Theoretical Physics and GDS thanks the Theory Unit at
CERN for their respective hospitality.  This work made extensive use of the
\healpix{} package~\citep{healpix}.  The numerical simulations were
performed on the facilities provided by the Case ITS High Performance
Computing Cluster.

\bibliographystyle{mn2e_new}
\bibliography{planck_r1_alignments}


\bsp

\label{lastpage}

\end{document}